\documentclass
[superscriptaddress,secnumarabic,amssymb,amsmath,nobibnotes,aps,prd,showkeys,showpacs,nofootinbib,twocolumn]{revtex4}%
\usepackage{graphicx}
\usepackage{subfigure}
\usepackage{epstopdf}
\usepackage{epsf}
\usepackage{bm}
\usepackage{amsmath}
\usepackage{amsfonts}
\usepackage{amssymb}%
\providecommand{\U}[1]{\protect\rule{.1in}{.1in}}

\newcommand{\be}{\begin{equation}}
\newcommand{\ee}{\end{equation}}

\newcommand{\mincir}{\raise
-3.truept\hbox{\rlap{\hbox{$\sim$}}\raise4.truept\hbox{$<$}\ }}
\newcommand{\magcir}{\raise
-3.truept\hbox{\rlap{\hbox{$\sim$}}\raise4.truept\hbox{$>$}\ }}

\begin{document}
\title{Inhomogeneous spacetimes in Weyl integrable geometry with matter source}
\author{Andronikos Paliathanasis}
\email{anpaliat@phys.uoa.gr}
\affiliation{Institute of Systems Science, Durban University of Technology, Durban 4000,
South Africa}
\affiliation{Instituto de Ciencias F\'{\i}sicas y Matem\'{a}ticas, Universidad Austral de
Chile, Valdivia 5090000, Chile}
\author{Genly Leon}
\email{genly.leon@ucn.cl}
\affiliation{Departamento de Matem\'{a}ticas, Universidad Cat\'{o}lica del Norte, Avda.
Angamos 0610, Casilla 1280 Antofagasta, Chile.}
\author{John D. Barrow}
\email{jdb34@hermes.cam.ac.uk}
\affiliation{DAMTP, Centre for Mathematical Sciences, University of Cambridge, Wilberforce
Rd., Cambridge CB3 0WA, UK}

\begin{abstract}
We investigate the existence of inhomogeneous exact solutions in Weyl
Integrable theory with a matter source. In particular, we consider the
existence of a dust fluid source while for the underlying geometry we assume a
line element which belongs to the family of silent universes. We solve
explicitly the field equations and we find the Szekeres spacetimes in Weyl
Integrable theory. We show that only the isotropic family can describe
inhomogeneous solutions where the LTB spacetimes are included. A detailed
analysis of the dynamics of the field equations is given where the past and
future attractors are determined. It is interesting that the Kasner spacetimes
can be seen as past attractors for the gravitation models, while the unique
future attractor describes the Milne universe similar with the behaviour of
the gravitational model in the case of General Relativity.

\end{abstract}
\keywords{Inhomogeneous spacetimes; Weyl theory; Exact solutions; Szekeres universes.}
\pacs{98.80.-k, 95.35.+d, 95.36.+x}
\date{\today}
\maketitle

\section{Introduction}

\label{sec1}

Analytical and exact solutions play a significant role in the study of
gravitational physics.\ The existence of exact spacetimes is essential in
order to understand the physical properties and the nature of the physical
space. Inhomogeneous and anisotropic exact spacetimes that have zero magnetic
Weyl tensor are very useful in gravitation and cosmology. They include an
important family of spacetimes known as the Szekeres universes. The Szekeres
spacetimes are the most general cosmological exact solutions of general
relativity with a pressureless fluid source \cite{szek0, kras}. They possess
no symmetries but the spatial three-slices have a special geometrical
structure. In the Szekeres spacetimes, information does not propagate via
gravitational or sound waves, so, they are also known as 'silent'
universes \cite{silent1}.

Szekeres spacetimes are inhomogeneous universes that do not admit any vector
field isometry. Moreover, the rotation and acceleration of the fluid source
must be identically zero, and the pressure constant. In practice, this means
the only inhomogeneous matter sources allowed are dust, with or without a
cosmological constant. While, in general, the spacetimes are anisotropic --
which means that the shear is non-zero and the expansion rate is non-zero. The
inhomogeneous Szekeres spacetimes are classified into two families: the
inhomogeneous Kantowski-Sachs (-like) spacetimes and the inhomogeneous FLRW
(-like) spacetimes.

There are applications of Szekeres spacetimes in gravitational physics and
cosmology \cite{szek0,bsilk}. A complete description of the scalar polynomial
curvature singularities in both classes of Szekeres solution have been
established, and they are velocity-dominated. They have Newtonian counterparts
and contain no gravitational waves \cite{newt,bon}. In addition, the
asymptotic behavior in the distant future has been analyzed
\cite{Bonnorasym,Goode:1982pg}.

A more general gravitational collapse, known as quasi-spherical by using the
Szekeres spacetimes was studied in \cite{sz1}, where it was found that a
strong radial increase in the density, the\ fluid heralds the onset of a naked
singularity. The matter distribution in Szekeres spaces has a dipolar
character \cite{shawB}, while there is no gravitational radiation emission
from the inhomogeneous moving dust \cite{bon}, for other applications of
Szekeres spacetimes in gravitational physics we refer the reader to
\cite{sz2,sz3,sz4,sz5} and references therein. Tilted Szekeres models were
studied in \cite{sz6} where it was found that vorticity follows the congruence
of the fluid world lines. Recently, the frame rotation of the Szekeres
spacetimes which relates the cosmological solutions with the quasi-spherical
exact solutions was studied in \cite{sz7a}.

The quasi--spherical Szekeres dust solutions are a generalization of the
spherically symmetric Lema\^{\i}tre--Tolman--Bondi dust models where the
spherical shells of constant mass are not concentric. A coordinate-independent
analysis of the dynamics of the spherically symmetric
Lema\^{\i}tre--Tolman-Bondi cosmologies, emphasizing their relation to the
Friedmann Lema\^{\i}tre cosmologies was given in \cite{Wainwright:2009zz}. In
general, it was shown that ever-expanding Lema\^{\i}tre--Tolman-Bondi
cosmologies isotropize at late times, approaching the de Sitter universe, or
the Milne universe, depending on whether or not a cosmological constant is
present. For the analysis, a dimensionless scalar is introduced to represent
the ratio of the Weyl and Ricci curvatures. In all cases, there is a finite
limit at late times, its value determines the asymptotic spatial
inhomogeneities in various physical quantities. The
Lema\^{\i}tre-Tolman-Bondi cosmologies for which the initial singularity is
isotropic were also identified. The collapsing quasi-spherical Szekeres dust
solution, where an apparent horizon covers all shell-crossings that will
occur, can be considered as a model for the formation of a black hole. The
apparent horizon can be detected by a Cartan invariant \cite{Coley:2019ylo}.
In the former reference, solutions of this sort are reviewed together with
their spin coefficients and curvature scalars in the Newman-Penrose formalism.
The Cartan--Karlhede algorithm is used to generate the minimal set of extended
Cartan invariants. Cartan scalars are compared with the kinematic scalars
\cite{Wainwright:2005} and q-scalars \cite{Sussman:2011bp}, which are two
well-known sets of scalars used to characterize Szekeres solutions.

Inhomogeneous spacetimes can been seen as limits of FLRW spacetimes with
inhomogeneous perturbations, such a comparison between non-spherical Szekeres
spaces and the dynamics of cosmological perturbation theory was performed in
\cite{sz7}.\ Specifically, it was proved that the linearised Szekeres
evolution equations and their solutions fully coincide with the corresponding
equations of the linear cosmological perturbation theory and their solutions
in the isochronous comoving gauge. Moreover, the conservation of the curvature
perturbation holds for the appropriate linear approximation of the exact
Szekeres fluctuations in $\Lambda$-cosmology, while the different collapse
morphologies of Szekeres models yield different growth factors to those that
follow from the analysis of redshift space distortions.

There are various generalizations of the Szekeres exact solutions where
additional matter sources contribute to the gravitating matter \cite{kras}.
Indeed, the first generalization presented by Szafron in \cite{ss1}, where the
dust fluid source was replaced by a perfect fluid with non-zero pressure,
leading to the Szekeres-Szafron spacetimes. The cosmological constant term was
introduced by Barrow et al. \cite{ss2} where the inhomogeneous analogue of the
$\Lambda$CDM model was derived. Other kinds of matter source have been
introduced, such as heat flow, electromagnetic field, viscosity, and an aether
field in the context of Einstein-aether theory
\cite{ss3,ss4,ss5,ss6,ss7,ss8,ss10,ss11}.

In this work, we are interested in determining exact inhomogeneous spacetimes
in Weyl Integrable theory \cite{cur0}. A Weyl manifold is a conformal manifold
equipped with a connection which preserves the conformal structure and is
torsion-free. In Weyl Integrable theory the connection structure is related to
the Levi-Civita connection, to which it differs by a scalar field of the
conformal metric. Specifically, if $g_{\mu\nu}$ is a metric tensor with
Levi-Civita connection $\Gamma_{\mu\nu}^{\kappa}$, then in Weyl Integrable
theory the manifold is supported by the set $\left\{  g_{\mu\nu},\tilde
{\Gamma}_{\mu\nu}^{\kappa}\right\}  $ where $\tilde{\Gamma}_{\mu\nu}^{\kappa}$
is the Levi-Civita connection for the conformally related metric $\tilde
{g}_{\mu\nu}=\phi g_{\mu\nu}$, where $\phi$ is a scalar field. An important
characteristic of the Weyl Integrable theory is that it is in agreement with
current astronomical and other observations \cite{cur1}.

Physical consequences of Weyl invariant theories are discussed, e.g., in
\cite{Quiros:2000tq,Quiros:2014hua,Quiros:2018ryt}. In \cite{Avalos:2016unj},
it is discussed whether or not a general Weyl structure is a suitable
mathematical model of spacetime. In this regard, it was found that a Weyl
integrable spacetime is the most general structure suitable to model
spacetime. The well-posedness of the Cauchy problem for particular kinds of
geometric scalar-tensor theories of gravity, which are based on a Weyl
integrable spacetime, is given in \cite{Avalos:2018uvq}. In
\cite{Aguila:2014moa}, a formulation of general relativity on a
Weyl-integrable geometry which contains cosmological solutions, exhibiting
acceleration in the present cosmic expansion, is studied. The conditions for
accelerated expansion of the universe are derived there. A particular solution
for the Weyl scalar field describing a cosmological model for the present time
is obtained in concordance with the data-combination Planck + WP + BAO + SN.
In \cite{va2}, the evolution of 4-, 5- and 6-dimensional cosmological models
based on the integrable Weyl geometry are considered numerically both for
empty spacetime and for scalar field with non-minimal coupling with gravity.
In \cite{Villanueva:2018kem}, the motion of massless particles on the
background of a toroidal topological black hole is analyzed in the context of
conformal Weyl gravity. Null geodesics are found analytically in terms of the
Jacobi elliptic functions.

There are various exact solutions of the field equations in Weyl Integrable
theory. Vacuum cosmological models were studied in \cite{va1}, while higher-
or lower-dimensional gravitational models were studied in \cite{va2,va3,va4}.
In \cite{salim96}, the authors studied gravitational models in Weyl Integrable
theory with matter source, an electromagnetic field, and additional scalar
field. In these models an interaction between the scalar field of the Weyl
theory and the matter sources is introduced, when the field equations are
written in the covariant form using the tensor quantities of general
relativity. Inhomogeneous models in Weyl Integrable models were also studied
in \cite{va5,va6,va7,va8}, while spherically symmetric solutions can be found
in \cite{vaa9}. In addition in \cite{vaa9} the authors discuss the
similarities and the differences of the Weyl Integrable theory with the
Brans-Dicke theory.

The plan of this paper is as follows. In\ Section \ref{sec2} we present the
basic properties and definitions of the gravitational field equations in Weyl
Integrable theory. We rewrite the field equations in a way that is equivalent
in form to general relativity and show that a scalar field is introduced in
the field equations, and we discuss the case where an ideal gas contributes in
the gravitational model. For the underlying geometry, we consider the line
element which belong to the silent universe class, and describes the Szekeres
spacetimes in general relativity. Exact solutions of the field equations are
presented in Section 4. In Sections \ref{sec5} we perform a detailed analysis
of the field equations in order to understand the past and future evolution of
the cosmological solutions. Finally, in Section 6 we discuss our results and
draw our conclusions.

\section{Weyl Integrable gravity}

\label{sec2}

Weyl geometry is an extension of Riemannian geometry, specified by a metric
tensor $g_{\mu\nu}$ and a gauge vector field $\omega_{\mu}$. The covariant
derivative $\tilde{\nabla}_{\mu}$ is defined by the (Weyl) affine connection
$\tilde{\Gamma}_{\mu\nu}^{\kappa},$ with the property,
\begin{equation}
\tilde{\nabla}_{\kappa}g_{\mu\nu}=\omega_{\kappa}g_{\mu\nu}, \label{ww.01}%
\end{equation}
from which we infer that the gauge vector $\omega_{\mu}~$ plays a significant
role in the geometry. Specifically, the Weyl affine connection $\tilde{\Gamma
}_{\mu\nu}^{\kappa}$ is related to the Christoffel symbols~$\Gamma_{\mu\nu
}^{\kappa}$ of the metric tensor $g_{\mu\nu}$ as follows:%
\begin{equation}
\tilde{\Gamma}_{\mu\nu}^{\kappa}=\Gamma_{\mu\nu}^{\kappa}-\omega_{(\mu}%
\delta_{\nu)}^{\kappa}+\frac{1}{2}\omega^{\kappa}g_{\mu\nu}. \label{ww.02}%
\end{equation}

The curvature tensor in the Weyl geometry is defined as
\begin{equation}
\tilde{\nabla}_{\nu}\left(  \tilde{\nabla}_{\mu}u_{\kappa}\right)
-\tilde{\nabla}_{\mu}\left(  \tilde{\nabla}_{\nu}u_{\kappa}\right)  =\tilde
{R}_{\kappa\lambda\mu\nu}u^{\lambda} \label{ww.03}%
\end{equation}
where, using (\ref{ww.02}), we observe that in general $\tilde{R}%
_{\kappa\lambda\mu\nu}$ is not symmetric as it is in the case of Riemannian geometry.

In this work, we are interested in the case where $\tilde{R}_{\kappa
\lambda\mu\nu}$ has the same symmetric properties as in Riemannian geometry.
This is true when $\omega_{\mu}$ is a gradient vector, which means that there
exists a scalar $\phi$ such that $\omega_{\mu}=\phi_{,\mu}$. In addition, in
this case, length variations are integrable along a closed path. This specific
theory is known as Weyl Integrable geometry. Moreover, there exists a
conformal map which relates the metric tensor $g_{\mu\nu}$ of a Riemannian
space into that of Weyl integrable space, which means that a Weyl integrable
space is also conformally a Riemann space.

In Weyl integrable geometry, the Ricci tensor $\tilde{R}_{\mu\nu}$ is related
to the Riemannian Ricci tensor $R_{\mu\nu}$ by,%
\begin{align}
&\tilde{R}_{\mu\nu}=R_{\mu\nu}-\tilde{\nabla}_{\nu}\left(  \tilde{\nabla}_{\mu
}\phi\right)  -\frac{1}{2}\left(  \tilde{\nabla}_{\mu}\phi\right)  \left(
\tilde{\nabla}_{\nu}\phi\right) \nonumber \\
&   -\frac{1}{2}g_{\mu\nu}\left(  \frac{1}%
{\sqrt{-g}}\left(  g^{\mu\nu}\sqrt{-g}\phi\right)  _{,\mu\nu}-g^{\mu\nu
}\left(  \tilde{\nabla}_{\mu}\phi\right)  \left(  \tilde{\nabla}_{\nu}%
\phi\right)  \right)  , \label{ww.04}%
\end{align}
where the Ricci scalar, in a four-dimensional manifold, is written as
\cite{salim96}%
\begin{equation}
\tilde{R}=R-\frac{3}{\sqrt{-g}}\left(  g^{\mu\nu}\sqrt{-g}\phi\right)
_{,\mu\nu}+\frac{3}{2}\left(  \tilde{\nabla}_{\mu}\phi\right)  \left(
\tilde{\nabla}_{\nu}\phi\right)  . \label{ww.05}%
\end{equation}

\subsection{Gravitational Action Integral}

We define the simple gravitational action Integral, which includes the Weyl
Ricci scalar $\tilde{R}$ and the field $\phi_{;\mu}$, as
\begin{equation}
S_{W}=\int dx^{4}\sqrt{-g}\left(  \tilde{R}+\xi\left(  \tilde{\nabla}_{\nu
}\left(  \tilde{\nabla}_{\mu}\phi\right)  \right)  g^{\mu\nu}\right)  ,
\label{ww.06}%
\end{equation}
where $\xi$ is an arbitrary coupling constant. At this point, we remark that
\begin{align}
&\left(  \tilde{\nabla}_{\nu}\left(  \tilde{\nabla}_{\mu}\phi\right)  \right)
g^{\mu\nu}=\frac{1}{\sqrt{-g}}\left(  g^{\mu\nu}\sqrt{-g}\phi\right)
_{,\mu\nu} \nonumber \\
& -2g^{\mu\nu}\left(  \tilde{\nabla}_{\mu}\phi\right)  \left(
\tilde{\nabla}_{\nu}\phi\right)  . \label{ww.07}%
\end{align}

Variation with respect to the metric tensor of the action integral, $S_{W}$,
provides the gravitational field equations \cite{salim96},%
\begin{align}
&\tilde{G}_{\mu\nu}+\tilde{\nabla}_{\nu}\left(  \tilde{\nabla}_{\mu}%
\phi\right)  -\left(  2\xi-1\right)  \left(  \tilde{\nabla}_{\mu}\phi\right)
\left(  \tilde{\nabla}_{\nu}\phi\right) \nonumber \\
&   +\xi g_{\mu\nu}g^{\kappa\lambda
}\left(  \tilde{\nabla}_{\kappa}\phi\right)  \left(  \tilde{\nabla}_{\lambda
}\phi\right)  =0, \label{ww.08}%
\end{align}
where $\tilde{G}_{\mu\nu}$ is the Weyl Einstein tensor. Moreover, variation
with respect to the scalar field $\phi$ gives%
\begin{equation}
\left(  \tilde{\nabla}_{\nu}\left(  \tilde{\nabla}_{\mu}\phi\right)  \right)
g^{\mu\nu}+2g^{\mu\nu}\left(  \tilde{\nabla}_{\mu}\phi\right)  \left(
\tilde{\nabla}_{\nu}\phi\right)  =0, \label{ww.09}%
\end{equation}
that is, a Klein-Gordon equation of the form,%
\begin{equation}
g^{\mu\nu}\nabla_{\nu}\nabla_{\mu}\phi=0, \label{ww.10}%
\end{equation}
where $\nabla_{\mu}$ denotes the Riemannian covariant derivative.

The gravitational field equations (\ref{ww.08}) can be rewritten  using the
Riemannian Einstein tensor $G_{\mu\nu}$ as follows%
\begin{equation}
G_{\mu\nu}-\lambda\left(  \phi_{,\mu}\phi_{,\nu}-\frac{1}{2}g_{\mu\nu}%
\phi^{,\kappa}\phi_{,\kappa}\right)  =0, \label{ww.11}%
\end{equation}
in which the new constant $\lambda$ is defined as $2\lambda\equiv4\xi-3$.

Consequently, the field equations (\ref{ww.11}) for $\lambda>0$ are those of
general relativity\footnote{We consider the signature of the metric to be
$\left(  -,+,+,+\right)  $.} with a massless scalar field. A new possibility
is introduced when $\lambda<0,$ which correspond to the addition of a massless
phantom scalar field.

Until now, we have considered the case of vacuum. Now, we present the field
equations in the presence of a matter source. Specifically, we consider the
cases where a pressureless dust fluid source contributes to the gravitational
field equations.

\subsection{The presence of dust}

When a pressureless fluid dust source is included, the gravitational field
equations become
\begin{equation}
G_{\mu\nu}-\lambda\left(  \phi_{,\mu}\phi_{,\nu}-\frac{1}{2}g_{\mu\nu}%
\phi^{,\kappa}\phi_{,\kappa}\right)  =T_{\mu\nu}^{\left(  m\right)  },
\label{ww.12}%
\end{equation}
where~$T_{\mu\nu}^{\left(  m\right)  }=e^{-\frac{\phi}{2}}\rho_{m}u_{\mu
}u_{\nu}$, while the Klein-Gordon equation (\ref{ww.10}) becomes
\begin{equation}
\frac{1}{\sqrt{-g}}\left(  g^{\mu\nu}\sqrt{-g}\phi_{,\mu}\right)  _{,\nu
}-\frac{1}{2\lambda}e^{-\frac{\phi}{2}}\rho_{m}=0, \label{ww.13}%
\end{equation}
while the conservation equation for the matter field reads $\tilde{\nabla
}_{\nu}T^{\left(  m\right)  \mu\nu}=0$.

We can see that there is a coupling between the scalar field and the dust
fluid; hence, the coupling and effective pressure term can depend on the
energy density $\rho_{m}$. The interaction between scalar field and dust fluid
has been proposed as a potential mechanism to explain the cosmic coincidence
problem. \cite{Amendola-ide1,Amendola-ide2,Pavon:2005yx,delCampo:2008sr}.
Various interaction models have been studied before in the literature; for
instance, see:
\cite{Amendola:2006dg,Pavon:2007gt,Chimento:2009hj,Arevalo:2011hh,Yang:2017zjs}%
, and references therein.

\section{Inhomogeneous spacetimes}

\label{sec3}

The gravitational model that we have considered in Weyl Integrable geometry is
equivalent to that of general relativity  with an effective energy-momentum
tensor, where the field equations are of the form%
\begin{equation}
G_{\mu\nu}=T_{\mu\nu}, \label{ww.19}%
\end{equation}
and $T_{\mu\nu}~$is the effective energy-momentum tensor. It consists of a
massless scalar field $\phi$, and an additional fluid source interacting with
the field $\phi$. In particular, $T_{\mu\nu}=T_{\mu\nu}^{\left(  \phi\right)
}+\hat{T}_{\mu\nu}$, where $\hat{T}_{\mu\nu}$ describes the energy-momentum
tensor of a pressureless fluid,~i.e.~$\hat{T}_{\mu\nu}=T_{\mu\nu}^{\left(
m\right)  },$ or of the second scalar field, $\psi$;$~$that is,
\begin{equation}
T_{\mu\nu}^{\left(  \psi\right)  }=e^{-2\phi}\left(  \psi_{,\mu}\psi_{,\nu
}-\frac{1}{2}g_{\mu\nu}\psi^{,\kappa}\psi_{,\kappa}-g_{\mu\nu}U\left(
\psi\right)  \right)  , \label{ww.20}%
\end{equation}
where $T_{\mu\nu}^{\left(  \phi\right)  }$ is the energy-momentum tensor of
the massless scalar field,
\begin{equation}
T_{\mu\nu}^{\left(  \phi\right)  }=\lambda\left(  \phi_{,\mu}\phi_{,\nu}%
-\frac{1}{2}g_{\mu\nu}\phi^{,\kappa}\phi_{,\kappa}\right)  . \label{ww.21}%
\end{equation}
However, as we discussed before the continuity equation $\nabla_{\nu}T^{\mu
\nu}=0$, provides~$\nabla_{\nu}\left(  T^{\left(  \phi\right)  \mu\nu}+\hat
{T}^{\mu\nu}\right)  =0$, that is, $\nabla_{\nu}\left(  T^{\left(
\phi\right)  \mu\nu}\right)  =Q,~\nabla_{\nu}\left(  \hat{T}^{\mu\nu}\right)
=-Q$, where $Q=Q\left(  x^{\mu}\right)  $, is the interacting term.

In this work, we assume that the underlying spacetime is described by the
inhomogeneous and anisotropic diagonal line element%
\begin{equation}
ds^{2}=-dt^{2}+e^{2A}dr^{2}+e^{2B}\left(  dy^{2}+dz^{2}\right)  ,
\label{ww.22}%
\end{equation}
in which $A=A\left(  t,r,y,z\right)  $ and $B=B\left(  t,r,y,z\right)  $. The
functional forms of the two scale factors $A$,~$B$ are determined by the
solution of the field equations (\ref{ww.19}).

In the case of general relativity, for $B_{,r}=0$, i.e. $B=B\left(
t,y,z\right)  $, the exact solutions belong to the Kantowski-Sachs (-like)
family, while for $B_{,r}\neq0$ the resulting spacetimes are inhomogeneous and
isotropic. \ Generalization of the Szekeres spacetimes with a purely
time-dependent scalar field have been studied before in \cite{ss9}. In
particular, a quintessence scalar field was considered and the scalar field
should be homogeneous. Although the density can be inhomogeneous in metric
(\ref{ww.22}), the pressure must be homogeneous \cite{kras}.

Hence, from the results for the vacuum solution of \cite{ss9} when the scalar
field is massless, we recover the exact solution of the vacuum Weyl Integrable
geometry for $\lambda>0$. However, in the presence of an additional matter
source, as we have here because of the existence of the interaction term, the
analytic solutions will be different.

In the following section, we proceed with the presentation of the analytic
solutions for the field equations (\ref{ww.19}), where the underlying
spacetime is described by the line element (\ref{ww.22}).

\label{sec4}

We require the pressure term of the effective energy-momentum tensor
$T_{\mu\nu}$ in (\ref{ww.19}) to be homogeneous such that the FLRW limit to be
provided. Thus, $\phi=\phi\left(  t\right)  $ while for the matter source we
have $\rho_{m}=\rho_{m}\left(  t\right)  $. The latter follows easily, if we
rewrite the energy-momentum tensor $T_{\mu\nu}$ such that to define a new
pressure component in order to eliminate the interaction term. The steps that
we follow to solve the field equations are similar to those taken in ref.
(\cite{ss9}). Thus, we omit the presentation and go directly to the main results.

Similarly, in the case of the homogeneous scalar field, we find that the
Szekeres-like solutions in the Weyl Integrable theory are classified into two
classes of solutions, (A) the inhomogeneous Kantowski-Sachs family of
solutions and the (B) inhomogeneous FLRW (-like) solutions.

For the Kantowski-Sachs family of solutions the unknown functions in the line
element (\ref{ww.22}) are $A\left(  t,r,y,z\right)  =\alpha\left(  t\right)  $
and $B\left(  t,r,y,z\right)  =\beta\left(  t\right)  \left(  c_{1}%
uv+c_{2}u+c_{3}v+c_{4}\right)  ,$ where the $y=u+v,~z=i\left(  u-v\right)  $,
so that the line element is written as \cite{ss9}:
\begin{equation}
ds^{2}=-dt^{2}+\alpha^{2}\left(  t\right)  dr^{2}+\beta^{2}\left(  t\right)
e^{2C\left(  y,z\right)  }\left(  dy^{2}+dz^{2}\right)  , \label{ww.23}%
\end{equation}
and the curvature, $K,~$of the two-dimensional surface of constant curvature
$\left\{  y-z\right\}  $ to be related with the constants $c_{1},$
$c_{2},~c_{3}$ and $c_{4}$ as follows, $K=c_{1}c_{4}-c_{2}c_{3}$. The unknown
time-dependent functions $\alpha\left(  t\right)  ,\beta\left(  t\right)  $
are determined by a set of differential equations that will be presented in
the following sections.

The second family of solutions which correspond to the inhomogeneous FLRW-like
spacetimes are described by the line element \cite{ss9}:%
\begin{align}
& ds^{2}=-dt^{2}+\alpha^{2}\left(  t\right)  \Bigg(  \left(  \frac{\partial
C\left(  r,y,z\right)  }{\partial r}\right)  ^{2}dr^{2} \nonumber \\
& +e^{2C\left(
r,y,z\right)  }\left(  dy^{2}+dz^{2}\right)  \Bigg)  . \label{ww.24}%
\end{align}
The function $C\left(  r,y,z\right)  $ is now given by~$C\left(  y,z\right)
=-2\ln\left(  \gamma_{1}\left(  r\right)  uv+\gamma_{2}\left(  r\right)
u+\gamma_{3}\left(  r\right)  v+\gamma_{4}\left(  r\right)  \right)  ,~$where
the functions $\gamma_{1}\left(  r\right)  ,~\gamma_{2}\left(  r\right)
,~\gamma_{3}\left(  r\right)  ,$ and $\gamma_{4}\left(  r\right)  $ are
constrained by $k=\gamma_{1}\left(  r\right)  \gamma_{4}\left(  r\right)
-\gamma_{2}\left(  r\right)  \gamma_{4}\left(  r\right)  $, where $k$ is the
spatial curvature of the FLRW-like spacetime. The scale factor $\alpha\left(
t\right)  $ is given by the generalized Friedmann equations in Weyl Integrable
geometry given below.

\subsection{Kantowski-Sachs spacetimes}

The unknown scale factors of the Kantowski-Sachs spacetime (\ref{ww.23}) are
given by the following system,%

\begin{equation}
\frac{2}{\alpha\beta}\dot{\alpha}\dot{\beta}+\frac{1}{\beta^{2}}\dot{\beta
}^{2}+\frac{K}{\beta^{2}}+\frac{\lambda}{2}\dot{\phi}^{2}+e^{-\frac{\phi}{2}%
}\rho_{m}=0, \label{ww.25}%
\end{equation}%
\begin{equation}
\frac{\ddot{\alpha}}{\alpha}+\frac{\ddot{\beta}}{\beta}+\frac{1}{\alpha\beta
}\dot{\alpha}\dot{\beta}+\frac{\lambda}{2}\dot{\phi}^{2}=0, \label{ww.26}%
\end{equation}%
\begin{equation}
2\frac{\ddot{\beta}}{\beta^{2}}+\frac{\dot{\beta}^{2}}{\beta^{2}}-\frac
{K}{\beta^{2}}+\frac{\lambda}{2}\dot{\phi}^{2}=0, \label{ww.27}%
\end{equation}
while the equation of motion for the scalar field and the matter source are
given by,%
\begin{equation}
\ddot{\phi}+\left(  \frac{\dot{\alpha}}{\alpha}+2\frac{\dot{\beta}}{\beta
}\right)  \dot{\phi}+\frac{1}{2\lambda}e^{-\frac{\phi}{2}}\rho_{m}=0,
\label{ww.28}%
\end{equation}%
\begin{equation}
\dot{\rho}_{m}+\left(  \frac{\dot{\alpha}}{\alpha}+2\frac{\dot{\beta}}{\beta
}-\dot{\phi}\right)  \rho_{m}=0, \label{ww.28a}%
\end{equation}
where overdot means total derivative with respect to the variable $t$.

\subsection{FLRW spacetimes}

Analogously, the unique scale factor for the FLRW (-like) spacetime
(\ref{ww.24}) is given by the (modified) Friedmann equations%
\begin{equation}
-3\left(  \frac{\dot{\alpha}}{\alpha}\right)  ^{2}+3k\alpha^{-2}+\frac
{\lambda}{2}\dot{\phi}^{2}+e^{-\frac{\phi}{2}}\rho_{m}=0, \label{ww.29}%
\end{equation}%
\begin{equation}
-2\frac{\ddot{\alpha}}{\alpha}-\left(  \frac{\dot{\alpha}}{\alpha}\right)
^{2}+k\alpha^{-2}-\frac{\lambda}{2}\dot{\phi}^{2}=0, \label{ww.30}%
\end{equation}
and the scalar field $\phi$ satisfies the Klein-Gordon equation,
\begin{equation}
\ddot{\phi}+3\frac{\dot{\alpha}}{\alpha}\dot{\phi}+\frac{1}{2\lambda}%
e^{-\frac{\phi}{2}}\rho_{m}=0, \label{ww.31}%
\end{equation}
while the conservation equation for the dust fluid source is%
\begin{equation}
\dot{\rho}_{m}+\left(  3\frac{\dot{\alpha}}{\alpha}-\dot{\phi}\right)
\rho_{m}=0. \label{ww.31a}%
\end{equation}

At this point, we remark that for $\lambda=0$, only the vacuum solutions of
general relativity are recovered, while the Szekeres spacetimes are recovered
when $\phi=\phi_{0}$ and $\lambda\rightarrow\infty$.~This is reminiscent of
the range of the constant Brans-Dicke parameter, $\omega$, in scalar-tensor
theory such, where the limit of general relativity to be recovered as
$\omega\rightarrow\infty$ \cite{farbd}. This family of spacetimes includes
also the inhomogeneous Lema\^{\i}tre-Tolman-Bondi (LTB) spacetimes \cite{LTB}.

In the following we show the analytic solution for the inhomogeneous FLRW
(-like) spacetime.

\subsection{Inhomogeneous analytic solution}

Now let us consider the case when the spatial curvature is zero, i.e. $k=0$.
The gravitational field equations can be rewritten in an equivalent form,%
\begin{align}
2\dot{H}+3H^{2}+\frac{\lambda}{2}\Phi^{2}  &  =0,\label{ww.32}\\
\dot{\Phi}+3H\Phi+\frac{3}{2\lambda}H^{2}-\frac{1}{4}\Phi^{2}  &  =0,
\label{ww.33}%
\end{align}
where $H=\frac{\dot{a}}{a}$ is the Hubble function and $\Phi=\dot{\phi}$.

We continue by defining the new variables $\left\{  R,\Theta\right\}  $ which
are given by\ the point transformation
\begin{equation}
H\equiv R\cos\Theta~,~\Phi\equiv\sqrt{\frac{6}{\lambda}}R\sin\Theta.
\label{ww.33a}%
\end{equation}

Therefore, the field equations (\ref{ww.32}), (\ref{ww.33}) in the new
coordinates are,
\begin{align}
& -4\sqrt{\frac{\lambda}{6}}\frac{\dot{R}}{R^{2}}=3\sqrt{\frac{6}{\lambda}}%
\cos\Theta\left(  3-2\cos^{2}\Theta\right) \nonumber\\
&   +\sin\Theta\left(  2\cos^{2}%
\Theta-1\right)  , \label{ww.34}%
\end{align}%
\begin{align}
& -4\sqrt{\frac{\lambda}{6}}\frac{\dot{\Theta}}{R}=\cos\Theta\left(  2\cos
^{2}\Theta-1\right) \nonumber \\
&  +\sqrt{6\lambda}\sin\Theta\left(  2\cos^{2}%
\Theta-1\right)  , \label{ww.35}%
\end{align}
from which it follows that the general algebraic solution expressed in
parametric form is%
\begin{align}
I_{0}  &  =-\frac{\left(  6\lambda-1\right)  }{2}R^{2} \nonumber \\
& +\left(  \sqrt{6\lambda
}-6\lambda\right)  \ln\left(  \sin\left(  \Theta\right)  -\cos\left(
\Theta\right)  \right)  \nonumber\\
&  -\left(  \sqrt{6\lambda}+6\lambda\right)  \ln\left(  \sin\left(
\Theta\right)  +\cos\left(  \Theta\right)  \right) \nonumber \\
&  +\left(  6\lambda
+1\right)  \ln\left(  6\sqrt{\lambda}\sin\Theta+\sqrt{6}\cos\Theta\right)  ,
\label{ww.36}%
\end{align}

where $I_{0}$ is constant. In the special case where $6\lambda=1$, the generic
algebraic solution follows%
\begin{equation}
I_{0}=-\frac{1}{2}R^{2}-\frac{1}{1+\tan\Theta}-\ln\left(  \sin^{2}\Theta
-\cos^{2}\Theta\right)  . \label{ww.37}%
\end{equation}

We continue our analysis by studying the dynamics of the field equations,
specifically, the ones of the (Weyl) Szekeres system.

\begin{figure*}[ptb]
\centering\includegraphics[width=0.7\textwidth]{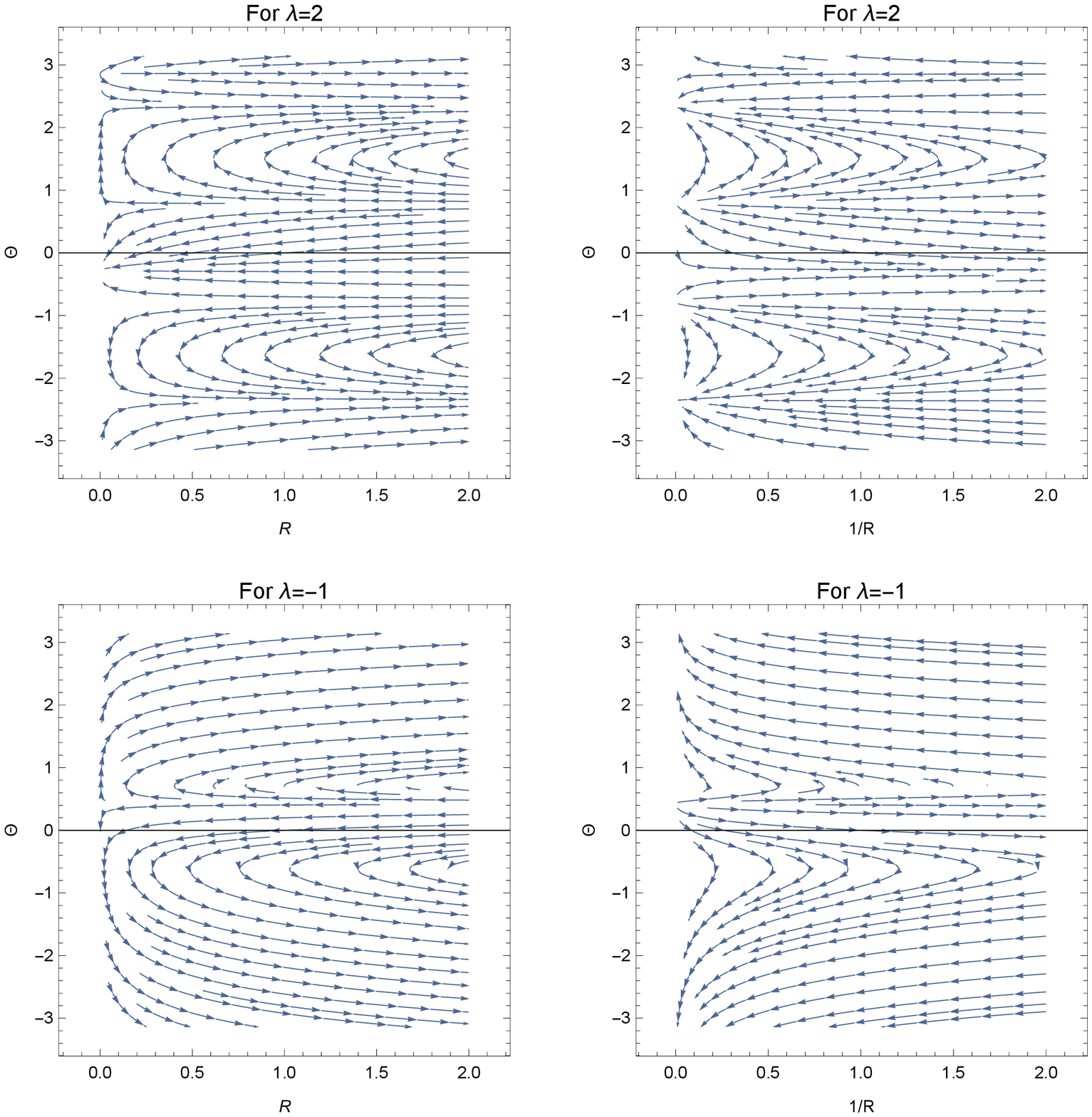} \caption{The phase
portrait of the dynamical system (\ref{ww.34}), (\ref{ww.35}) for $\lambda=2$
and $\lambda=-1$. Note that for negative values of $\lambda,$ we apply the
transformation $\Theta\rightarrow i\Theta$. }%
\label{pp1}%
\end{figure*}

The phase space portrait of the field equations (\ref{ww.34}), (\ref{ww.35})
is presented in Fig. \ref{pp1}.

\section{Dynamical analysis}

\label{sec5}

The field equations (\ref{ww.19}) with time-derivatives can be written in a
covariant form  using the kinematic variables for the observer: the volume
expansion rate $\theta=3H,~$, the shear scalar $\sigma$, the electric part of
the Weyl tensor $\mathit{E}$, and the components of the effective fluid energy
density $\rho$ and pressure $p$.

In particular, the field equations are then expressed as follows
\cite{Wainwright:2005,ellis2}%
\begin{subequations}
\begin{align}
\dot{\rho}+\theta\left(  \rho+p\right)   &  =0,~\label{ss.01}\\
\dot{\theta}+\frac{\theta^{2}}{3}+6\sigma^{2}+\frac{1}{2}\left(
\rho+3p\right)   &  =0,\label{ss.02}\\
\dot{\sigma}-\sigma^{2}+\frac{2}{3}\theta\sigma+\mathit{E}  &
=0,\label{ss.04}\\
\mathit{\dot{E}}+3\mathit{E}\sigma+\theta\mathit{E}+\frac{1}{2}\left(
\rho+p\right)  \sigma &  =0, \label{ss.05}%
\end{align}
with the constraint equation,
\begin{equation}
\frac{\theta^{2}}{3}-3\sigma^{2}+\frac{^{\left(  3\right)  }R}{2}=\rho,
\label{ss.06}%
\end{equation}
where $^{\left(  3\right)  }R$ is the spatial curvature of the
three-dimensional hypersurfaces. The latter system is known as the
Szekeres-Szafron system and has been widely studied in the literature
\cite{silent1,sil2,sil3,sil4}.

In Weyl Integrable theory with a dust fluid source the effective energy
density and pressure are $\rho=e^{-\frac{\phi}{2}}\rho_{m}+\rho_{\phi
}~,~p=p_{\phi},~$in which~$\rho_{\phi}=\frac{\lambda}{2}\dot{\phi}^{2}%
~$and$~p_{\phi}=\frac{\lambda}{2}\dot{\phi}^{2}$. \ In addition, from equation
(\ref{ss.01}) we can write the equivalent system%
\end{subequations}
\begin{subequations}
\begin{equation}
\dot{\rho}_{m}+\left(  \theta-\dot{\phi}\right)  \rho_{m}=0, \label{ww.51}%
\end{equation}%
\begin{equation}
\dot{\rho}_{\phi}+\theta\left(  \rho_{\phi}+p_{\phi}\right)  +\frac
{e^{-\frac{\phi}{2}}}{2}\rho_{m}\dot{\phi}=0. \label{ww.52}%
\end{equation}

In the following, we rewrite the field equations (\ref{ss.01})-(\ref{ss.06})
using expansion-normalized variables to determine the stationary points of
the dynamical system. We remark that every stationary point corresponds to an
exact solution of the field equations, which can describe a specific epoch
provided by the dynamics of the system. The stability of the stationary points
is also determined, which is needed to determine the past and future evolution
of the solutions provided by the stationary points. 

\subsection{Dimensionless variables}

We define the new expansion-normalised dimensionless variables%
\end{subequations}
\begin{equation}
\Omega_{m}=\frac{3e^{-\frac{\phi}{2}}\rho_{m}}{\theta^{2}},~\Omega_{R}%
=\frac{3R}{2\theta^{2}},x=\frac{\sqrt{6}\dot{\phi}}{2\theta},~\beta
=\frac{\sigma}{\theta},~\alpha=\frac{\mathit{E}}{\theta^{2}}.
\label{ww.57}%
\end{equation}
In the new variables, the Szekeres system becomes
\begin{subequations}
\begin{footnotesize}
\begin{align}
\Omega_{m}^{\prime}  &  =\frac{1}{2}\Omega_{m}\left(  \sqrt{6}x+8\lambda
x^{2}+72\beta^{2}+2\left(  \Omega_{m}-1\right)  \right)  ,\label{ww.58}\\
x^{\prime}  &  =\frac{1}{12\lambda}\left(  2\lambda x\left(  36\beta
^{2}+4\lambda x^{2}+\Omega_{m}-4\right)  -\sqrt{6}\Omega_{m}\right)
,\label{ww.61}\\
\beta^{\prime}  &  =\frac{1}{2}\left(  6\beta^{2}\left(  1+6\beta\right)
+\beta\Omega-2\left(  \beta+3\alpha\right)  +4\lambda\beta x^{2}\right)
,\label{ww.59}\\
\alpha^{\prime}  &  =\frac{1}{2}\left(  2\alpha\left(  \Omega+4\lambda
x^{2}-1+9\beta\left(  4\beta-1\right)  \right)  -\beta\left(  2\lambda
x^{2}+\Omega_{m}\right)  \right)  , \label{ww.60}%
\end{align}
\end{footnotesize}
with (first integral) constraint equation%
\begin{equation}
\Omega_{R}=-1+9\beta^{2}+\lambda x^{2}+\Omega_{m}, \label{ww.62}%
\end{equation}
\end{subequations}
where the prime derivative is defined by $\Omega_{m}^{\prime}\equiv
\frac{d\Omega_{m}}{d\tau}$,~where $\tau=\ln a$ and $a(\tau)$ is the geometric
mean expansion scale factor ($\dot{a}/a=H)$. Moreover, the parameter for the
equation of state of an effective fluid source, $w_{tot}=\frac{p}{\rho}$, is
expressed in terms of the dimensionless variables as

\begin{equation}
w_{tot}=\frac{1}{3}\left(  \Omega-1+4x^{2}\lambda+36\beta^{2}\right)  .
\label{ww.63}%
\end{equation}
\begin{figure*}[ptb]
\centering\includegraphics[width=0.7\textwidth]{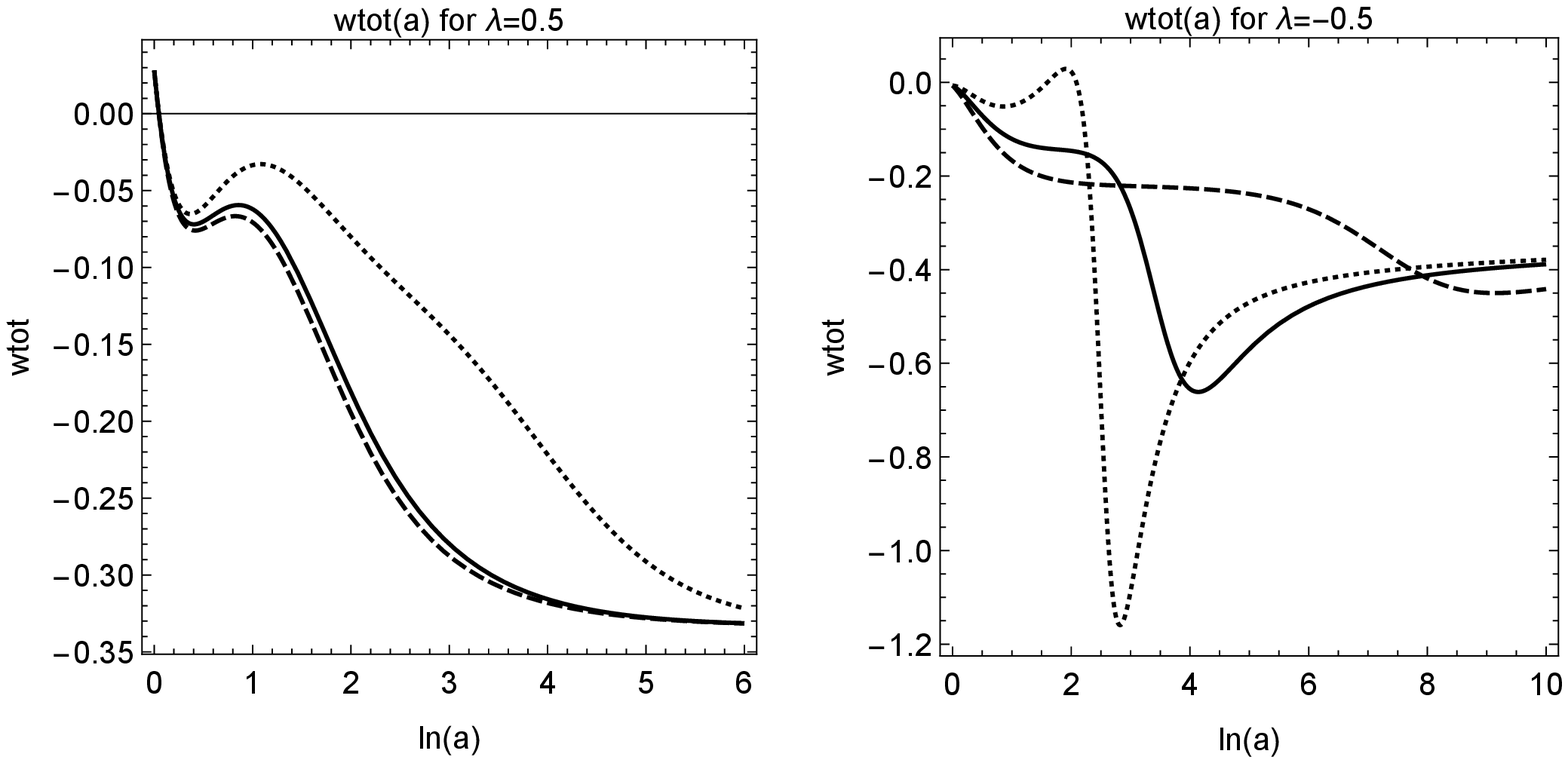}
\caption{Qualitative evolution of the parameter, $w$, for the equation of
state of the effective fluid, for various initial conditions. The left-hand
figure is for a positive value of $\lambda=0.5,$ while right-hand figure is
for a negative value of $\lambda=-0.5$. We observe that the future attractor
is the Milne universe, $w_{tot}\left(  P_{1}\right)  =-\frac{1}{3}$. The
left-hand figure is for initial conditions, $\Omega_{m0}=0.75,~\beta
_{0}=0.02,~x_{0}=0.4,$ $\alpha_{0}=0$ (solid line), $\alpha_{0}=0.01$ (dashed
line) and $\alpha_{0}=-0.01$ (dotted line). The right-hand figure is for
initial conditions, $\Omega_{m0}=0.75,~\beta_{0}=0.08,~x_{0}=0.02,$
$\alpha_{0}=0$ (solid line), $\alpha_{0}=0.01$ (dashed line) and $\alpha
_{0}=-0.01$ (dotted line).}%
\label{dust1}%
\end{figure*}

\subsection{Stationary points}

The set of stationary points,~$\mathbf{P,}$ have coordinates$~\mathbf{P}%
=\left(  \Omega_{m}\left(  \mathbf{P}\right)  ,x\left(  \mathbf{P}\right)
,\beta\left(  \mathbf{P}\right)  ,\alpha\left(  \mathbf{P}\right)  \right)  $,
and the physical properties of the exact solutions at these points for the
four-dimensional dynamical system (\ref{ww.58})-(\ref{ww.61}) are presented below.%

\begin{table*}[tbp] \centering
\caption{Stationary points and their stability for the Szekeres system in Weyl Integrable geometry with a dust fluid source.}%
\begin{tabular}
[c]{ccccc}\hline\hline
\textbf{Point} & $\left(  {\Omega}_{m},\mathbf{x},{\beta},{\alpha}\right)  $ &
${\Omega}_{R}$ & Spacetime & Stability\\\hline
$P_{1}$ & $\left(  0,0,0,0\right)  $ & $-1$ & FLRW\ (Milne Universe) &
Stable\\
$P_{2}$ & $\left(  0,0,\frac{1}{6},0\right)  $ & $-\frac{3}{4}$ &
Kantowski-Sachs & Unstable\\
$P_{3}$ & $\left(  0,0,-\frac{1}{3},0\right)  $ & $0$ & Bianchi I & Unstable\\
$P_{4}$ & $\left(  0,0,\frac{1}{3},\frac{2}{9}\right)  $ & $0$ & Bianchi I &
Unstable\\
$P_{5}$ & $\left(  0,0,-\frac{1}{12},\frac{1}{32}\right)  $ & $-\frac{15}{6}$
& Kantowski-Sachs & Unstable\\
$P_{6}$ & $\left(  0,x,\pm\sqrt{1-\lambda x^{2}},\frac{1}{9}\left(  1-\lambda
x^{2}\pm\sqrt{1-\lambda x^{2}}\right)  \right)  $ & $0$ & Bianchi I &
Unstable\\
$P_{7}$ & $\left(  1-\frac{1}{6\lambda},-\frac{1}{\sqrt{6}\lambda},0,0\right)
$ & $0$ & FLRW (spatially flat) & Unstable\\
$P_{8}$ & $\left(  -\frac{8}{3}\lambda,\sqrt{\frac{2}{3}},0,0\right)  $ &
$-1-2\lambda$ & FLRW (open) & Unstable\\
$P_{9}$ & $\left(  -\frac{3\lambda\left(  2\lambda+5\right)  }{2\left(
\lambda+2\right)  ^{2}},\frac{1}{2+\lambda}\sqrt{\frac{3}{2}},-\frac{1}%
{3}+\frac{1}{4+2\lambda},\frac{3+4\lambda\left(  \lambda+2\right)  }{24\left(
\lambda+2\right)  ^{2}}\right)  $ & $-\frac{3\left(  2\lambda+1\right)
\left(  2\lambda+5\right)  }{4\left(  \lambda+2\right)  ^{2}}$ &
Kantowski-Sachs & Unstable\\\hline\hline
\end{tabular}
\label{tab1}%
\end{table*}%

Point $P_{1}=\left(  0,0,0,0\right)  $ describes an empty isotropic universe,
with spatial curvature $\Omega_{R}\left(  P_{1}\right)  =-1$ and a parameter
for the equation of state $w_{tot}\left(  P_{1}\right)  =-\frac{1}{3}$. From
the latter, we infer that the exact solution at the point $P_{1}$ is the Milne
universe. In order to infer the stability of the exact solution at point
$P_{1}$, we determine the eigenvalues of the linearized system around $P_{1}$.
They are $e_{1}\left(  P_{1}\right)  =-1$,$~e_{2}\left(  P_{1}\right)  =-1$,
$e_{3}\left(  P_{1}\right)  =-1$ and $e_{4}\left(  P_{1}\right)  =-\frac{2}%
{3}$, hence $P_{1}$ is an attractor.

Point $P_{2}=\left(  0,0,\frac{1}{6},0\right)  $ has physical quantities
$\Omega_{R}\left(  P_{2}\right)  =-\frac{3}{4}$. This point describes an
anisotropic Kantowski-Sachs universe. The eigenvalues of the linearized system
are $e_{1}\left(  P_{2}\right)  =0$, $e_{2}\left(  P_{2}\right)  =-\frac{3}%
{4}$, $e_{3}\left(  P_{2}\right)  =-\frac{1}{2}$ and $e_{4}\left(
P_{2}\right)  =\frac{3}{2}$, from which we infer that $P_{2}$ is a saddle
point, that is, the exact solution at this point is unstable.

Point $P_{3}=\left(  0,0,-\frac{1}{3},0\right)  $ describes a vacuum Bianchi I
universe, $\Omega_{R}\left(  P_{3}\right)  =0,$ and more specifically, the
Kasner universe. The eigenvalues of the linearized system are $e_{1}\left(
P_{3}\right)  =0$,$~e_{2}\left(  P_{3}\right)  =3$, $e_{3}\left(
P_{3}\right)  =3$ and $e_{4}\left(  P_{3}\right)  =6$, hence $P_{3}$ is a source.

Point $P_{4}=\left(  0,0,\frac{1}{3},\frac{2}{9}\right)  $ describes a vacuum
Kasner universe, $\Omega_{R}\left(  P_{4}\right)  =0$, and the exact solution
is unstable. The eigenvalues are $e_{1}\left(  P_{4}\right)  =0$%
,$~e_{2}\left(  P_{4}\right)  =2$, $e_{3}\left(  P_{4}\right)  =3$ and
$e_{4}\left(  P_{4}\right)  =5$.

Point $P_{5}=\left(  0,0,-\frac{1}{12},\frac{1}{32}\right)  $ gives
$\Omega_{R}\left(  P_{5}\right)  =-\frac{15}{6}$ which means that the exact
solution at the point describes a Kantowski-Sachs universe. The eigenvalues of
the linearized system are $e_{1}\left(  P_{5}\right)  =-\frac{15}{8}$%
,$~e_{2}\left(  P_{5}\right)  =-\frac{3}{4}$, $e_{3}\left(  P_{5}\right)
=-\frac{5}{8}$ and $e_{4}\left(  P_{5}\right)  =\frac{3}{4}$, which means that
$P_{5}$ is a saddle point.

Points $P_{6}^{\pm}=\left(  0,x,\pm\sqrt{1-\lambda x^{2}},\frac{1}{9}\left(
1-\lambda x^{2}\pm\sqrt{1-\lambda x^{2}}\right)  \right)  $ are surfaces in
the phase space where $\Omega_{R}\left(  P_{6}\right)  =0$, which means that
the points describe Bianchi I spacetimes. The points are real when
$1-x^{2}\lambda\geq0$. In the limit where $x^{2}=\frac{1}{\lambda}$ the
solution reduces to that of isotropic FLRW spacetime with a stiff fluid
source. The eigenvalues of the linearized system are $e_{1}\left(
P_{6}\right)  =0,~e_{2}\left(  P_{6}\right)  =\frac{6+\sqrt{6}x}{2},~$%
\[
e_{3}\left(  P_{6}^{\pm}\right)  =4-\frac{4}{3}\lambda x^{2}\mp\frac
{\sqrt{1-\lambda x^{2}}}{2}\]\[+\frac{1}{6}\sqrt{81+\lambda x^{2}\left(  64\lambda
x^{2}-81-48\sqrt{1-\lambda x^{2}}\right)  },
\]%
\[
e_{4}\left(  P_{6}^{\pm}\right)  =4-\frac{4}{3}\lambda x^{2}\mp\frac
{\sqrt{1-\lambda x^{2}}}{2}\]\[-\frac{1}{6}\sqrt{81+\lambda x^{2}\left(  64\lambda
x^{2}-81-48\sqrt{1-\lambda x^{2}}\right)  }.
\]
Point $P_{7}=\left(  1-\frac{1}{6\lambda},-\frac{1}{\sqrt{6}\lambda
},0,0\right)  $, describes a FLRW spacetime, $\Omega_{R}\left(  P_{7}\right)
=0$, where the equation of state parameter for the effective fluid is
$w_{tot}\left(  P_{7}\right)  =\frac{1}{6\lambda}$. This point is physically
acceptable when $\lambda>\frac{1}{6}$, which means that $0<w_{tot}\left(
P_{7}\right)  <1$. The eigenvalues of the linearized system are $e_{1}\left(
P_{7}\right)  =1+\frac{1}{2\lambda}$,$~e_{2}\left(  P_{7}\right)  =-\frac
{3}{2}+\frac{1}{4\lambda}$, $e_{3}\left(  P_{7}\right)  =\frac{1}{4}+\frac
{5}{72\lambda}+\frac{\sqrt{2916\lambda^{2}+1044\lambda-191}}{72\lambda}%
$,~$e_{4}\left(  P_{7}\right)  =\frac{1}{4}+\frac{5}{72\lambda}-\frac
{\sqrt{2916\lambda^{2}+1044\lambda-191}}{72\lambda}$. Whence it follows
that the exact solution at point $P_{7}$ is unstable.

Point $P_{8}=\left(  -\frac{8}{3}\lambda,\sqrt{\frac{2}{3}},0,0\right)  $ is
physical acceptable for $-\frac{3}{8}<\lambda<0\ $. It describes a FLRW
spacetime with spatial curvature $\Omega_{R}\left(  P_{8}\right)
=-1-2\lambda,$ which is always negative for the accepted values of $\lambda$.
The eigenvalues of the linearized system are derived to be,
$e_{1}\left(  P_{8}\right)  =-1+i\sqrt{2\lambda}$,$~e_{2}\left(  P_{8}\right)
=-1-i\sqrt{2\lambda}$, $e_{3}\left(  P_{8}\right)  =-\frac{1}{3}-\frac{8}%
{9}\lambda+\frac{1}{9}\left(  \sqrt{64\lambda^{2}+156\lambda+63}\right)
$,~$e_{4}\left(  P_{8}\right)  =-\frac{1}{3}-\frac{8}{9}\lambda-\frac{1}%
{9}\left(  \sqrt{64\lambda^{2}+156\lambda+63}\right)  $ from which we conclude
that the exact solution at $P_{8}$ is always unstable.

Point $P_{9}=\left(  -\frac{3\lambda\left(  2\lambda+5\right)  }{2\left(
\lambda+2\right)  ^{2}},\frac{1}{2+\lambda}\sqrt{\frac{3}{2}},-\frac{1}%
{3}+\frac{1}{4+2\lambda},\frac{3+4\lambda\left(  \lambda+2\right)  }{24\left(
\lambda+2\right)  ^{2}}\right)  $ describes a Kantowski-Sachs universe where
$\Omega_{R}\left(  P_{9}\right)  =-\frac{3\left(  2\lambda+1\right)  \left(
2\lambda+5\right)  }{4\left(  \lambda+2\right)  ^{2}}$. The point is physical
acceptable for $-\frac{23-\sqrt{273}}{16}\leq\lambda<0$ and $-\frac{5}{2}%
\leq\lambda<-\frac{23+\sqrt{273}}{16}$. The eigenvalues are calculated
numerically, from which we infer that point $P_{9}$ is a saddle point.

\begin{figure*}[ptb]
\centering\includegraphics[width=0.8\textwidth]{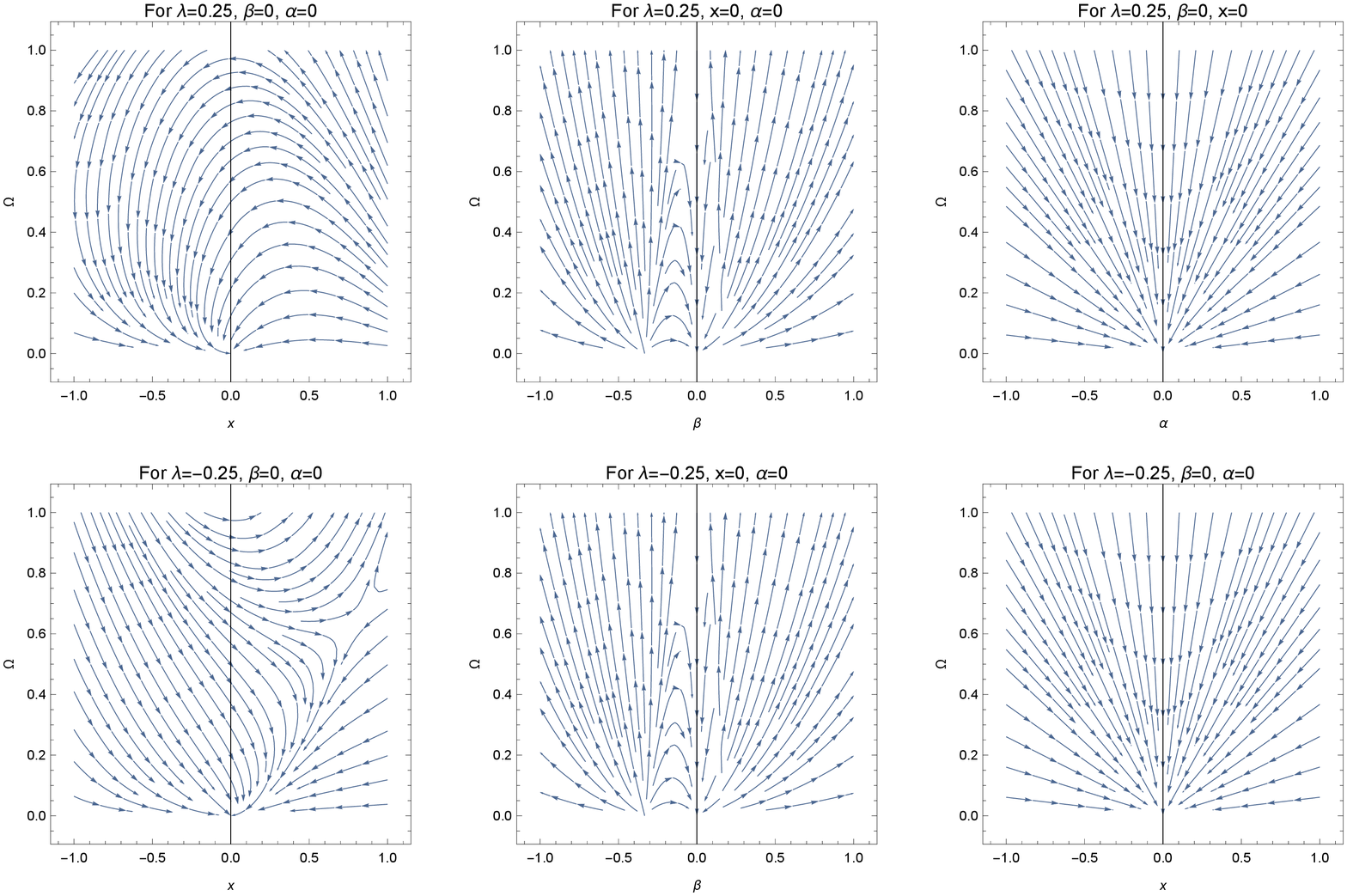}
\caption{Two-dimensional phase space portraits in the planes $\left\{
x-\Omega\right\}  $,~$\left\{  \beta-\Omega\right\}  $ and $\left\{
\alpha-\Omega\right\}  $. The figures of the first row are for $\lambda=0.25$,
while the figures of the second row are for $\lambda=-0.25$. The unique
attractor of the system is the Milne Universe.}%
\label{dust2}%
\end{figure*}

\begin{figure*}[ptb]
\centering\includegraphics[width=0.8\textwidth]{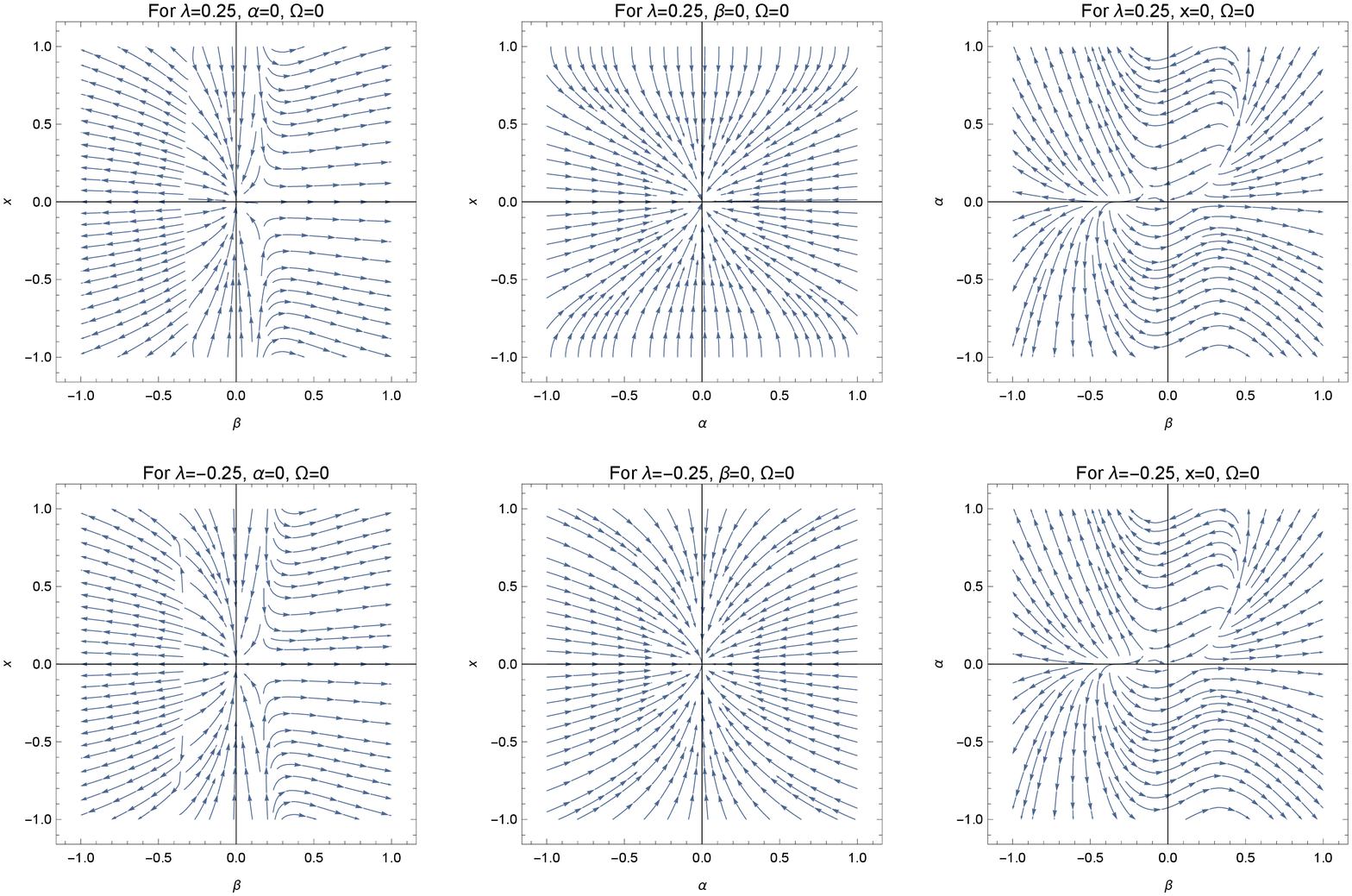}
\caption{Two-dimensional phase space portraits in the planes $\left\{
\beta-x\right\}  $,~$\left\{  \alpha-x\right\}  $ and $\left\{  \beta
-\alpha\right\}  $. The figures in the first row are for $\lambda=0.25$, while
the figures in the second row are for $\lambda=-0.25$. The unique attractor of
the system is the Milne Universe.}%
\label{dust3}%
\end{figure*}

The above results are summarized in Table \ref{tab1}. In Fig. \ref{dust1}, the
qualitative behaviour of the equation of state parameter $w_{tot}$ is
presented. Moreover, two-dimensional phase portraits for the dynamical system
(\ref{ww.58})-(\ref{ww.61}) are presented in Figs. \ref{dust2} and \ref{dust3}
where $P_{1}$ is the unique attractor. The plots are for positive and negative
values of the coupling parameter $\lambda$.

\subsection{Past attractors}

When analyzing the dynamics of the system (\ref{ww.58})-(\ref{ww.60}) towards
the past, it is convenient to make a time reversal $\tau\mapsto-\tau$. In this
case, we have the same points as before, but there is an overall change of
sign in the eigenvalues. Then, the possible late-time attractors of the new
system, given by
\begin{subequations}
\begin{footnotesize}
\label{EQ:52}%
\begin{align}
\Omega_{m}^{\prime}  &  =-\frac{1}{2}\Omega_{m}\left(  \sqrt{6}x+8\lambda
x^{2}+72\beta^{2}+2\left(  \Omega_{m}-1\right)  \right)  ,\\
x^{\prime}  &  =-\frac{1}{12\lambda}\left(  2\lambda x\left(  36\beta
^{2}+4\lambda x^{2}+\Omega_{m}-4\right)  -\sqrt{6}\Omega_{m}\right)  ,\\
\beta^{\prime}  &  =-\frac{1}{2}\left(  6\beta^{2}\left(  1+6\beta\right)
+\beta\Omega-2\left(  \beta+3\alpha\right)  +4\lambda\beta x^{2}\right)  ,\\
\alpha^{\prime}  &=-\frac{1}{2}\left(  2\alpha\left(  \Omega+4\lambda
x^{2}-1+9\beta\left(  4\beta-1\right)  \right)  -\beta\left(  2\lambda
x^{2}+\Omega_{m}\right)  \right),
\end{align}
\end{footnotesize}
\end{subequations}
correspond to the past attractors of the original one. We study the points
$P_{3}$ with coordinates $(\Omega_{m},x,\beta,\alpha)=\left(  0,0,-\frac{1}%
{3},0\right)  $, $\Omega_{R}=0$, and $P_{4}$ with coordinates $(\Omega
_{m},x,\beta,\alpha)=\left(  0,0,\frac{1}{3},\frac{2}{9}\right)  $,
$\Omega_{R}=0$, corresponding to Bianchi I models, and we show they are
unstable for the original system using the center manifold theorem (CMT). The
detailed analysis of the CMT for these two points is presented in Appendix
\ref{append1}.

\section{Conclusions}

In this work we found exact inhomogeneous spacetimes which generalize the
Szekeres universes into the Weyl integrable theory. Specifically, we assume
that the scalar field which defines the Weyl affine connection to be
homogeneous such that the limit of FLRW exists. In such scenario, the only
inhomogeneous spacetimes are those which belong to the FLRW (-like) solutions
included in the family of LTB spacetimes. On the other hand, the Kantowski-Sachs
family of solutions is homogeneous and anisotropic. For the inhomogeneous
spacetimes, we were able to write in terms of quadratics the generic solution
of the field equations.

In order to understand the dynamics and the evolution of the gravitational
model we performed a detailed study of the past and future attractors. In
particular, we defined Hubble-normalized dimensionless variables.\ The field
equations admit three stationary points which describe a spatially flat, an
open, and a closed FLRW space where only the closed FLRW spacetime can be a
future attractor, which gives the Milne universe. The other two isotropic
solutions correspond to saddle points.

In addition, three homogeneous Kantowski-Sachs spacetimes are supported by the
field equations which correspond to saddle points. There are three points
which describe Bianchi I spacetimes; the exact solution at one of these points
describes a Bianchi I spacetime with a stiff fluid, while the other two points
describe vacuum Kasner solutions. The points which describe the Kasner
solutions are sources while the third point is a saddle point.

For the sources we performed a detailed study on the past-system in order to
investigate if the points are attractors for the past-system. Indeed with the
application of the center manifold theorem we were able to prove that the
Kasner solutions are past attractors for the field equations for $\lambda>0$.

\begin{acknowledgments}
AP \& GL were funded by Agencia Nacional de Investigaci\'{o}n y Desarrollo -
ANID through the program FONDECYT Iniciaci\'{o}n grant no. 11180126.
Additionally, GL is supported by Vicerrector\'{\i}a de Investigaci\'{o}n y
Desarrollo Tecnol\'{o}gico at Universidad Catolica del Norte. Ellen de los
Milagros Fernández Flores is acknowledged for proofreading. JDB
is supported by the STFC of the United Kingdom.
\end{acknowledgments}

\appendix

\section{CMT for points $P_{3}$ and $P_{4}$}

\label{append1}

\subsection{Analysis of $P_{3}$}

Introducing the coordinate transformations
\begin{footnotesize}
\begin{align}
&\alpha\mapsto v_{3}-v_{1},\beta\mapsto v_{1}+\sqrt{\frac{2}{3}}\lambda
v_{2}-\frac{1}{3}, \nonumber\\
& \Omega_{m}\mapsto-6\sqrt{6}\lambda v_{2},x\mapsto u+v_{2},
\end{align}
\end{footnotesize}
the equilibrium point $P_{3}$ is translated to the origin and the
linearization matrix is transformed to its canonical real Jordan form.

Therefore, we obtain the equivalent dynamical system to \eqref{EQ:52}, defined
by%
\begin{small}
\begin{align*}
u^{\prime}  &  =-\frac{1}{3}\lambda(u-5v_{2})\left(  2u^{2}+4uv_{2}%
+v_{2}\left(  \sqrt{6}(12v_{1}-7)+2v_{2}\right)  \right)  \\
&   -6v_{1}^{2}%
(u-5v_{2})+4v_{1}(u-5v_{2})\\
&  -4\lambda^{2}v_{2}^{2}(u-5v_{2})+\sqrt{\frac{3}{2}}v_{2}(u+v_{2}),
\\
v_{1}^{\prime}  &  =\frac{1}{3}\Big\{-3v_{1}\left(  \lambda\left(  2\left(
(u+v_{2})^{2}-6\lambda v_{2}^{2}\right)  -5\sqrt{6}v_{2}\right)  +6\right) \\
& 
+2\lambda^{2}v_{2}\left(  \sqrt{6}(u+v_{2})^{2}-18v_{2}\right) \\
&  +\lambda\left(  (u+v_{2})(2u+5v_{2})-3\sqrt{6}v_{2}\right)  -54v_{1}%
^{3} \\
& +9v_{1}^{2}\left(  5-2\sqrt{6}\lambda v_{2}\right)  +12\sqrt{6}\lambda
^{3}v_{2}^{3}+9v_{3}\Big\},
\\
& v_{2}^{\prime}=-\frac{1}{2}v_{2}\Big(  8\lambda u^{2}+u\left(  16\lambda
v_{2}+\sqrt{6}\right)  +72v_{1}^{2} \\
& +48v_{1}\left(  \sqrt{6}\lambda
v_{2}-1\right)  +8\lambda(6\lambda+1)v_{2}^{2}+\sqrt{6}(1-28\lambda
)v_{2}+6\Big),
\\
v_{3}^{\prime}  &  =3v_{1}\left(  \lambda\left(  (u+v_{2})^{2}+12\lambda
v_{2}^{2}-5\sqrt{6}v_{2}\right)  +v_{3}\left(  11-8\sqrt{6}\lambda
v_{2}\right)  \right)  \\
& +\lambda^{2}v_{2}\left(  \sqrt{6}\left(  (u+v_{2}%
)^{2}+4\lambda v_{2}^{2}\right)  -18v_{2}\right) \\
&  -v_{3}\left(  \lambda\left(  4\left(  (u+v_{2})^{2}+6\lambda v_{2}%
^{2}\right)  -17\sqrt{6}v_{2}\right)  +3\right)  \\
& +\frac{1}{3}\lambda
(u+v_{2})(u+4v_{2})+18v_{1}^{3}-18v_{1}^{2}\left(  -\sqrt{6}\lambda
v_{2}+2v_{3}+1\right).
\end{align*}
\end{small}
The eigen system of the origin is
\[
\left(
\begin{array}
[c]{cccc}%
-6 & -3 & -3 & 0\\
\{0,1,0,0\} & \{0,1,0,1\} & \left\{  0,-\sqrt{\frac{2}{3}}\lambda,1,0\right\}
& \{1,0,0,0\}\\
&  &  &
\end{array}
\right)  .
\]
That is, the center manifold of the origin is tangent to the $u$-axis, and it
is given locally by a graph
\begin{small}
\begin{align}
&\Big\{  (u,v_{1},v_{2},v_{3})\in\mathbb{R}^{4}:  v_{i}=h_{i}(u), \nonumber \\
& h_{i}%
(0)=h_{i}^{\prime}(0)=0,i=1\ldots4,|u|<\delta\Big\}  ,
\end{align}
\end{small}
which satisfies the differential equations
\begin{small}
\begin{align}
F(u,h_{1},h_{2},h_{3})h_{1}^{\prime}(u)+G_{1}(u,h_{1},h_{2},h_{3})  &
=0,\label{s1}\\
2F(u,h_{1},h_{2},h_{3})h_{2}^{\prime}(u)+G_{2}(u,h_{1},h_{2},h_{3})  &
=0,\label{s2}\\
2F(u,h_{1},h_{2},h_{3})h_{3}^{\prime}(u)+G_{3}(u,h_{1},h_{2},h_{3})  &  =0,
\label{s3}%
\end{align}
\end{small}
where 
\begin{footnotesize}
\begin{align*}
&F(u,h_{1},h_{2},h_{3})    =12h_{1}(u-5h_{2})\left(  \sqrt{6}\lambda
h_{2}-1\right) \\
&  +18h_{1}^{2}(u-5h_{2})+ \lambda(u-5h_{2})\left(  h_{2}\left(  2h_{2}+4u-7\sqrt{6}\right)
+2u^{2}\right) \\ & +12\lambda^{2}(u-5h_{2})h_{2}^{2}-3\sqrt{\frac{3}{2}}%
h_{2}(h_{2}+u),
\\
&G_{1}(u,h_{1},h_{2},h_{3})  = \\& -3h_{1}\left(  \lambda h_{2}\left(
(2-12\lambda)h_{2}+4u-5\sqrt{6}\right)  +2\lambda u^{2}+6\right) +\\
& +9h_{1}%
^{2}\left(  5-2\sqrt{6}\lambda h_{2}\right)    -54h_{1}^{3} \\
& +\lambda h_{2}\Big( h_{2}\left(  2\sqrt{6}(6\lambda+1)\lambda
h_{2}+4\lambda\left(  \sqrt{6}u-9\right)  +5\right) \\ 
&    +u\left(  2\sqrt
{6}\lambda u+7\right)  -3\sqrt{6}\Bigg) +9h_{3}+2\lambda u^{2},
\end{align*}%
\begin{align*}
& G_{2}(u,h_{1},h_{2},h_{3})=-3h_{2}\Big(  48h_{1}\left(  \sqrt{6}\lambda
h_{2}-1\right)  +72h_{1}^{2} \\ 
& +h_{2}\left(  8(6\lambda+1)\lambda h_{2}%
-28\sqrt{6}\lambda+16\lambda u+\sqrt{6}\right)  +u\left(  8\lambda u+\sqrt
{6}\right)  +6\Big),
\\
 & G_{3}(u,h_{1},h_{2},h_{3})  =h_{1}\Big(  h_{2}\left(  \lambda\left(
36u-90\sqrt{6}\right)  -144\sqrt{6}\lambda h_{3}\right) \\
&  +18\lambda
(12\lambda+1)h_{2}^{2}+198h_{3}+18\lambda u^{2}\Big)  +\\
&  +h_{1}^{2}\left(  108\sqrt{6}\lambda h_{2}-216h_{3}-108\right)
+108h_{1}^{3} \\
& +h_{2}\left(  6\lambda\left(  17\sqrt{6}-8u\right)
h_{3}+2\lambda u\left(  3\sqrt{6}\lambda u+5\right)  \right)  +\\
&  +h_{2}^{2}\left(  4\lambda\left(  3\lambda\left(  \sqrt{6}u-9\right)
+2\right)  -24\lambda(6\lambda+1)h_{3}\right) \\
& +6\sqrt{6}\lambda^{2}%
(4\lambda+1)h_{2}^{3}-6h_{3}\left(  4\lambda u^{2}+3\right)  +2\lambda u^{2}.%
\end{align*}
\end{footnotesize}
Using Taylor expansions, we propose as Ans\"{a}tze: 
\begin{widetext}
\begin{small}
\[
\left(
\begin{array}
[c]{c}%
h_{1}(u)\\
h_{2}(u)\\
h_{3}(u)\\
\end{array}
\right)  =\left(
\begin{array}
[c]{c}%
a_{1}u^{2}+a_{2}u^{3}+a_{3}u^{4}+a_{4}u^{5}+a_{5}u^{6}+a_{6}u^{7}+a_{7}%
u^{8}+a_{8}u^{9}+a_{9}u^{10}+a_{10}u^{11}+a_{11}u^{12}+a_{12}u^{13}+\ldots\\
b_{1}u^{2}+b_{2}u^{3}+b_{3}u^{4}+b_{4}u^{5}+b_{5}u^{6}+b_{6}u^{7}+b_{7}%
u^{8}+b_{8}u^{9}+b_{9}u^{10}+b_{10}u^{11}+b_{11}u^{12}+b_{12}u^{13}+\ldots\\
c_{1}u^{2}+c_{2}u^{3}+c_{3}u^{4}+c_{4}u^{5}+c_{5}u^{6}+c_{6}u^{7}+c_{7}%
u^{8}+c_{8}u^{9}+c_{9}u^{10}+c_{10}u^{11}+c_{11}u^{12}+c_{12}u^{13}+\ldots\\
\end{array}
\right).
\]
\end{small}
\end{widetext}
Substituting in (\ref{s1})-(\ref{s3}), and equating the coefficients of equal
powers of $u$, we obtain%
\begin{align*}
a_{1}  &  =\frac{\lambda}{6},a_{2}=0,a_{3}=\frac{\lambda^{2}}{24}%
,a_{4}=0,a_{5}=\frac{\lambda^{3}}{48},a_{6}=0, \\
& a_{7}=\frac{5\lambda^{4}}%
{384},a_{8}=0,a_{9}=\frac{7\lambda^{5}}{768},a_{10}=0,a_{11}=\frac
{7\lambda^{6}}{1024},a_{12}=0,\\
b_{1}  &  =0,b_{2}=0,b_{3}=0,b_{4}=0,b_{5}=0,b_{6}=0, \\ & b_{7}=0,b_{8}%
=0,b_{9}=0,b_{10}=0,b_{11}=0,b_{12}=0,\\
c_{1}  &  =\frac{\lambda}{9},c_{2}=0,c_{3}=\frac{\lambda^{2}}{18}%
,c_{4}=0,c_{5}=\frac{\lambda^{3}}{36},c_{6}=0, \\ & c_{7}=\frac{5\lambda^{4}}%
{288},c_{8}=0,c_{9}=\frac{7\lambda^{5}}{576},c_{10}=0,c_{11}=\frac
{7\lambda^{6}}{768},c_{12}=0.
\end{align*}
Therefore,%
\begin{align*}
\alpha &  \mapsto-\frac{\lambda u^{2}}{18}+\frac{\lambda^{2}u^{4}}{72}%
+\frac{\lambda^{3}u^{6}}{144} \\ & +\frac{5\lambda^{4}u^{8}}{1152}+\frac
{7\lambda^{5}u^{10}}{2304}+\frac{7\lambda^{6}u^{12}}{3072}+\ldots,\\
\beta &  \mapsto-\frac{1}{3}+\frac{\lambda u^{2}}{6} +\frac{\lambda^{2}u^{4}%
}{24}+\frac{\lambda^{3}u^{6}}{48}  \\ & +\frac{5\lambda^{4}u^{8}}{384}+\frac
{7\lambda^{5}u^{10}}{768}+\frac{7\lambda^{6}u^{12}}{1024}+\ldots,\\
\Omega_{m}  &  \mapsto0,\\
x  &  \mapsto u,
\end{align*}
and we have the parametrization,%
\begin{align*}
\dot{\phi}  &  =\sqrt{\frac{2}{3}}\theta(u+v_{2})\sim\sqrt{\frac{2}{3}}\theta
u+\mathcal{O}(u)^{14},\\
\rho_{m}  &  =-2\sqrt{6}\theta^{2}\lambda v_{2}e^{\frac{\phi}{2}}%
\sim\mathcal{O}(u)^{14},
\end{align*}
\begin{align*}
\sigma &  =\frac{1}{3}\theta\left(  3v_{1}+\sqrt{6}\lambda v_{2}-1\right) \\
& 
\sim\theta\Bigg(  \frac{7\lambda^{6}u^{12}}{1024}+\frac{7\lambda^{5}u^{10}%
}{768}+\frac{5\lambda^{4}u^{8}}{384} \\
& +\frac{\lambda^{3}u^{6}}{48}+\frac
{\lambda^{2}u^{4}}{24}+\frac{\lambda u^{2}}{6}-\frac{1}{3}\Bigg)
+\mathcal{O}(u)^{14},\\
\mathit{E}  &  =\theta^{2}(v_{3}-v_{1})\sim \\
& \theta^{2}\lambda\Bigg(
\frac{7\lambda^{5}u^{12}}{3072}+\frac{7\lambda^{4}u^{10}}{2304}+\frac
{5\lambda^{3}u^{8}}{1152} \\
& +\frac{\lambda^{2}u^{6}}{144}+\frac{\lambda u^{4}%
}{72}-\frac{u^{2}}{18}\Bigg)  +\mathcal{O}(u)^{14},
\end{align*}
where we choose $\lambda v_{2}\geq0$.

The dynamics on the center manifold of the origin are dictated by a
gradient-like equation $u^{\prime}=-\nabla U(u)$. For $\lambda>0$,
$\omega=u\sqrt{\lambda}$, the equation transforms to
\begin{equation}
\scriptscriptstyle 
\omega^{\prime}=-\frac{\omega^{15}\left(  \left(  7\left(  63\omega
^{6}+168\omega^{4}+352\omega^{2}+704\right)  \omega^{2}+10560\right)
\omega^{2}+33792\right)  }{1572864},
\end{equation}
for which the origin is a degenerated minimum.

For $\lambda<0$, $\omega=u\sqrt{-\lambda}$, the equation transforms to
\begin{equation}
\scriptscriptstyle \omega^{\prime}=\frac{\omega^{15}\left(  33792-\omega^{2}\left(
10560-7\omega^{2}\left(  704-\omega^{2}\left(  352-21\omega^{2}\left(
8-3\omega^{2}\right)  \right)  \right)  \right)  \right)  }{1572864},
\end{equation}
for which the origin is a degenerated maximum.

Therefore, for $\lambda>0$ (respectively, $\lambda<0$) the center manifold,
and hence, the origin of the dynamical system  is a local attractor
(respectively, a saddle). In the original variables this means that for
$\lambda>0$ the past attractor of system is $P_{3}$, and for $\lambda<0$ it is
a saddle point.

Now, we take the time reversal back and work in terms of $t$. We deduce:%
\begin{align*}
\dot{\theta}  &  =\theta^{2}\Bigg(  -\frac{147\lambda^{12}u^{24}}{524288}
-\frac{49\lambda^{11}u^{22}}{65536}-\frac{77\lambda^{10}u^{20}}{49152} \\ & 
-\frac{77\lambda^{9}u^{18}}{24576}-\frac{55\lambda^{8}u^{16}}{8192}
-\frac{11\lambda^{7}u^{14}}{512}-1\Bigg) \\
&  \sim-\theta^{2}-\frac{11}{512}\left(  \theta^{2}\lambda^{7}\right)
u^{14}+\mathcal{O}\left(  u^{16}\right)  ,
\end{align*}%
\begin{align*}
\dot{u}  &  =\frac{1}{3}\theta\Bigg(  \frac{147\lambda^{12}u^{25}}{524288}
+\frac{49\lambda^{11}u^{23}}{65536}+\frac{77\lambda^{10}u^{21}}{49152} \\ & 
+\frac{77\lambda^{9}u^{19}}{24576}+\frac{55\lambda^{8}u^{17}}{8192}
+\frac{11\lambda^{7}u^{15}}{512}\Bigg) \\
&  \sim\frac{11\left(  \theta\lambda^{7}\right)  u^{15}}{1536}+\mathcal{O}%
\left(  u^{16}\right)  .
\end{align*}
The solutions can be expressed as:
\begin{footnotesize}
\begin{equation}
\theta(t)=\frac{1}{t-t_{0}}+\varepsilon c_{1}(t),\quad u(t)=\frac{\varepsilon
c_{2}(t)}{\sqrt[14]{-\ln(t-t_{0})}},\varepsilon\ll1,
\end{equation}
where%
\begin{align*}
c_{1}^{\prime}(t)  &  =-\frac{2c_{1}(t)}{t-t_{0}}-\varepsilon c_{1}(t){}%
^{2}+\mathcal{O}(\varepsilon^{13}),\\
c_{2}^{\prime}(t)  &  =\frac{c_{2}(t)}{14(t-t_{0})\ln(t-t_{0})}+\mathcal{O}%
(\varepsilon^{13}).
\end{align*}
Then,
\begin{equation}
c_{1}(t)=\frac{1}{(t-t_{0})\left(  c_{3}(t-t_{0})-\varepsilon\right)  },\quad
c_{2}(t)=c_{4}\sqrt[14]{-\ln(t-t_{0})}.
\end{equation}
Finally,
\begin{subequations}
\begin{align}
& \theta(t)=-\frac{c_{3}}{c_{3}(t_{0}-t)+\varepsilon}\nonumber \\
& \sim\frac{1}{t-t_{0}}%
+\frac{\varepsilon}{c_{3}(t-t_{0})^{2}}+\frac{\varepsilon^{2}}{(t-t_{0}%
)^{3}c_{3}^{2}}+\mathcal{O}\left(  \varepsilon^{3}\right), \\
& u=c_{4} \varepsilon,
\end{align}
and%
\end{subequations}
\begin{align}
\sigma &  =-\frac{1}{3(t-t_{0})}-\frac{\varepsilon}{3\left(  (t-t_{0}%
)^{2}c_{3}\right)  }+ \nonumber \\
& \frac{\left(  (t-t_{0})^{2}\lambda c_{4}^{2}-\frac
{2}{c_{3}^{2}}\right)  \varepsilon^{2}}{6(t-t_{0})^{3}}+\mathcal{O}\left(
\varepsilon^{3}\right)  ,\\
\mathit{E}  &  =-\frac{\left(  \lambda c_{4}^{2}\right)  \varepsilon^{2}%
}{18(t-t_{0})^{2}}+\mathcal{O}\left(  \varepsilon^{3}\right)  ,\\
\dot{\phi}  &  =\frac{\sqrt{\frac{2}{3}}c_{4}\varepsilon}{t-t_{0}}+\frac
{\sqrt{\frac{2}{3}}c_{4}\varepsilon^{2}}{(t-t_{0})^{2}c_{3}}+\mathcal{O}%
\left(  \varepsilon^{3}\right)  ,\\
\phi &  =\sqrt{\frac{2}{3}}c_{4}\varepsilon\ln(t-t_{0})-\frac{\sqrt{\frac
{2}{3}}c_{4}\varepsilon^{2}}{c_{3}(t-t_{0})}+\mathcal{O}\left(  \varepsilon
^{3}\right)  ,\\
\rho_{m}  &  =\frac{2\varepsilon^{14}\left(  \sqrt{6}c_{4}^{14}K_{0}\right)
}{(t-t_{0})^{2}}+O\left(  \varepsilon^{15}\right)  ,
\end{align}
\end{footnotesize}
where $c_{3}$ and $c_{4}$ are integration constants, and we set $\lambda
v_{2}=-K_{0}\varepsilon^{14}$, for a positive constant $K_{0}$. For
$\lambda>0$, $\theta(t)\rightarrow\frac{1}{t-t_{0}}$ as $t\rightarrow0$
($\tau\rightarrow-\infty$). Hence, $P_{3}$ it is associated with (an
anisotropic) initial singularity.

\subsection{Analysis of $P_{4}$}

Introducing the coordinate transformation
\begin{footnotesize}
\begin{align}
& \alpha\mapsto\frac{3v_{1}}{2}+\sqrt{6}\lambda v_{2}+\frac{3v_{3}}{5}+\frac
{2}{9}, \beta\mapsto v_{1}+\sqrt{6}\lambda v_{2}+v_{3}+\frac{1}{3}%
,\nonumber\\
& \Omega\mapsto-6\sqrt{6}\lambda v_{2},\quad x\mapsto u+v_{2},
\end{align}
\end{footnotesize}
the equilibrium point $P_{4}$ is translated to the origin and the
linearization matrix is transformed to its canonical real Jordan form.

Therefore, we obtain the equivalent dynamical system to \eqref{EQ:52}, defined
by%
\begin{small}
\begin{align*}
u^{\prime} &  =\frac{1}{6}\Big\{-2\lambda(u-5v_{2})\Big(  2u^{2}%
+4uv_{2} \\ & +v_{2}\left(  9\sqrt{6}(4v_{1}+4v_{3}+1)+2v_{2}\right)  \Big)  \\
&  +3\Big(  v_{2}\left(  20(v_{1}+v_{3})(3v_{1}+3v_{3}+2)+\sqrt{6}%
v_{2}\right) \\ &  -u\left(  12v_{1}^{2}+8v_{1}(3v_{3}+1)-\sqrt{6}v_{2}%
+4v_{3}(3v_{3}+2)\right)  \Big)  \\
&  -216\lambda^{2}v_{2}^{2}(u-5v_{2})\Big\},
\\
{v_{1}}^{\prime} &  =\frac{1}{405}\Big\{-15v_{1}\Big(  3\lambda\left(
38(u+v_{2})^{2}+3888\lambda v_{2}^{2}+145\sqrt{6}v_{2}\right) \\
 &  +v_{3}\left(
3888\sqrt{6}\lambda v_{2}+669\right)  +1944v_{3}^{2}+55\Big)  \nonumber\\
&  +6\Big[-15\lambda^{2}v_{2}\left(  \sqrt{6}\left(  (u+v_{2})^{2}+324\lambda
v_{2}^{2}\right)  +60v_{2}\right) \\
&  +v_{3}\left(  -15\lambda\left(
(u+v_{2})^{2}+972\lambda v_{2}^{2}+32\sqrt{6}v_{2}\right)  -16\right)
\nonumber\\
&  -30v_{3}^{2}\left(  81\sqrt{6}\lambda v_{2}+11\right)  -810v_{3}%
^{3}\Big] \\
& -10\lambda(7u-47v_{2})(u+v_{2})-19440v_{1}^{3}\nonumber\\
&  -45v_{1}^{2}\left(  972\sqrt{6}\lambda v_{2}+972v_{3}+179\right)  \Big\},
\\
{v_{2}}^{\prime} &  =\frac{1}{2}v_{2}\Big\{-u\left(  8\lambda u+\sqrt
{6}\right) \\
&  -v_{2}\left(  16\lambda u+\sqrt{6}(36\lambda(4v_{1}+4v_{3}%
+1)+1)\right)  \\
&  -6(2v_{1}+2v_{3}+1)(6v_{1}+6v_{3}+1)-8\lambda(54\lambda+1)v_{2}^{2}\Big\},
\\
{v_{3}}^{\prime} &  =\frac{1}{162}\Big\{3v_{1}\Big(  60\lambda\left(
2(u+v_{2})^{2}+486\lambda v_{2}^{2}+\sqrt{6}v_{2}\right) \\
& +6v_{3}\left(
972\sqrt{6}\lambda v_{2}-155\right)  +972v_{3}^{2}-25\Big)  \\
&  +360\lambda^{2}v_{2}\left(  \sqrt{6}\left(  (u+v_{2})^{2}+81\lambda
v_{2}^{2}\right)  +6v_{2}\right) \\
&  -6v_{3}\left(  3\lambda\left(
16(u+v_{2})^{2}-2916\lambda v_{2}^{2}+71\sqrt{6}v_{2}\right)  +134\right)  \\
&  -10\lambda(8u-19v_{2})(u+v_{2})+4860v_{1}^{3} \\
& +36v_{1}^{2}\left(
405\sqrt{6}\lambda v_{2}+243v_{3}-5\right) \\
&  +18v_{3}^{2}\left(  162\sqrt
{6}\lambda v_{2}-145\right)  -972v_{3}^{3}\Big\}.
\end{align*}
\end{small}
The eigen system of the origin is
\[
\left(
\begin{array}
[c]{cccc}%
-5 & -3 & -2 & 0\\
\left\{  0,\frac{2}{25},0,1\right\}   & \{0,0,1,0\} & \left\{  0,-\frac{32}%
{5},0,1\right\}   & \{1,0,0,0\}\\
&  &  &
\end{array}
\right)  .
\]
That is, the center manifold of the origin is tangent to the $u$-axis, and it
is given locally by a graph
\begin{small}
\begin{align}
&\Big\{  (u,v_{1},v_{2},v_{3})\in\mathbb{R}^{4}:  v_{i}=h_{i}(u), \nonumber \\
& h_{i}%
(0)=h_{i}^{\prime}(0)=0,i=1\ldots4,|u|<\delta\Big\}  ,
\end{align}
\end{small}
which satisfies the differential equations%
\begin{align}
F(u,h_{1},h_{2},h_{3})h_{1}^{\prime}(u)+G_{1}(u,h_{1},h_{2},h_{3}) &  =0,\\
F(u,h_{1},h_{2},h_{3})h_{2}^{\prime}(u)+G_{2}(u,h_{1},h_{2},h_{3}) &  =0,\\
F(u,h_{1},h_{2},h_{3})h_{3}^{\prime}(u)+G_{3}(u,h_{1},h_{2},h_{3}) &  =0,
\end{align}
where%
\begin{footnotesize}
\begin{align*}
& F(u,h_{1},h _{2},h_{3})  =h_{1}\Big(  h_{2}\left(  4\left(  3\sqrt{6}\lambda
u-5\right)  -60h_{3}\right) \\
&  -60\sqrt{6}\lambda h_{2}^{2}+12uh_{3}+4u\Big) \\
& +h_{1}^{2}(6u-30h_{2})  +h_{2}\Big(  4h_{3}\left(  3\sqrt{6}\lambda u-5\right) \\
&   -30h_{3}^{2}-2\lambda u^{2}+\sqrt{\frac{3}{2}}(6\lambda-1)u\Big)  +\\
&  +h_{2}^{2}\left(  -60\sqrt{6}\lambda h_{3}-\sqrt{\frac{3}{2}}%
(30\lambda+1)+6\lambda(6\lambda-1)u\right) \\
& -\frac{10}{3}\lambda
(54\lambda+1)h_{2}^{3}+6uh_{3}^{2}+4uh_{3}+\frac{2\lambda u^{3}}{3},
\\
& G_{1}(u,h_{1},h_{2},h_{3})  =h_{1}\Big( h_{2}\left(  -144\sqrt{6}\lambda
h_{3}-\frac{1}{9}\lambda\left(  76u+145\sqrt{6}\right)  \right) \\
& -\frac{2}{9}\lambda(1944\lambda+19)h_{2}^{2}-72h_{3}^{2}-\frac{223h_{3}}{9}%
-\frac{38\lambda u^{2}}{9}-\frac{55}{27}\Big)  +\\
&  +h_{1}^{2}\left(  -108\sqrt{6}\lambda h_{2}-108h_{3}-\frac{179}{9}\right)
-48h_{1}^{3} \\
& +h_{2}^{2}\left(  \frac{2}{81}\lambda\left(  47-18\lambda\left(
\sqrt{6}u+30\right)  \right)  -\frac{2}{9}\lambda(972\lambda+1)h_{3}\right)
+\end{align*}
\begin{align*}
&  +h_{2}\left(  -36\sqrt{6}\lambda h_{3}^{2}-\frac{4}{9}\lambda\left(
u+16\sqrt{6}\right)  h_{3}-\frac{2}{81}\lambda u\left(  9\sqrt{6}\lambda
u-40\right)  \right)  +\\
&  -\frac{2}{3}\sqrt{\frac{2}{3}}\lambda^{2}(324\lambda+1)h_{2}^{3}-\frac
{2}{135}h_{3}\left(  15\lambda u^{2}+16\right) \\
&  -12h_{3}^{3}-\frac{44h_{3}%
^{2}}{9}-\frac{14\lambda u^{2}}{81},
\end{align*}
\begin{align*}
& G_{2}(u,h_{1},h_{2},h_{3})  =h_{2}^{2}\Big(  -72\sqrt{6}\lambda
h_{1}-72\sqrt{6}\lambda h_{3} \\
& -\sqrt{\frac{3}{2}}(36\lambda+1)-8\lambda
u\Big) \\
&  +h_{2}\Big(  h_{1}(-72h_{3}-24)-36h_{1}^{2}-36h_{3}^{2}-24h_{3} \\
& +\frac{1}{2}\left(  -u\left(  8\lambda u+\sqrt{6}\right)  -6\right)  \Big)
-4\lambda(54\lambda+1)h_{2}^{3},
\\
 &  G_{3}(u,h_{1},h_{2},h_{3})=h_{1}\Big(  h_{2}\left(  108\sqrt{6}\lambda
h_{3}+\frac{10}{9}\lambda\left(  4u+\sqrt{6}\right)  \right) \\
&  +\frac{20}%
{9}\lambda(243\lambda+1)h_{2}^{2}+18h_{3}^{2}-\frac{155h_{3}}{9}+\frac{5}%
{54}\left(  24\lambda u^{2}-5\right)  \Big)  +\\
&  +h_{1}^{2}\left(  90\sqrt{6}\lambda h_{2}+54h_{3}-\frac{10}{9}\right)
+30h_{1}^{3} \\
& +h_{2}^{2}\left(  \frac{4}{9}\lambda(729\lambda-4)h_{3}+\frac
{5}{81}\lambda\left(  72\lambda\left(  \sqrt{6}u+3\right)  +19\right)
\right)  +\\
&  +h_{2}\Big(  18\sqrt{6}\lambda h_{3}^{2}-\frac{1}{9}\lambda\left(
32u+71\sqrt{6}\right)  h_{3} \\
& +\frac{5}{81}\lambda u\left(  36\sqrt{6}\lambda
u+11\right)  \Big)  +\\
&  +\frac{20}{3}\sqrt{\frac{2}{3}}\lambda^{2}(81\lambda+1)h_{2}^{3}-\frac
{2}{27}h_{3}\left(  24\lambda u^{2}+67\right) \\  
& -6h_{3}^{3}-\frac{145h_{3}^{2}}{9}-\frac{40\lambda u^{2}}{81}.
\end{align*}
\end{footnotesize}
Using Taylor expansion we propose as Ans\"{a}tze:
\begin{widetext}
 {\small
\begin{equation}
\left(
\begin{array}
[c]{c}%
h_{1}(u)\\
h_{2}(u)\\
h_{3}(u)\\
\end{array}
\right)  =\left(
\begin{array}
[c]{c}%
a_{1}u^{2}+a_{2}u^{3}+a_{3}u^{4}+a_{4}u^{5}+a_{5}u^{6}+a_{6}u^{7}+a_{7}%
u^{8}+a_{8}u^{9}+a_{9}u^{10}+a_{10}u^{11}+a_{11}u^{12}+a_{12}u^{13}+\ldots\\
b_{1}u^{2}+b_{2}u^{3}+b_{3}u^{4}+b_{4}u^{5}+b_{5}u^{6}+b_{6}u^{7}+b_{7}%
u^{8}+b_{8}u^{9}+b_{9}u^{10}+b_{10}u^{11}+b_{11}u^{12}+b_{12}u^{13}+\ldots\\
c_{1}u^{2}+c_{2}u^{3}+c_{3}u^{4}+c_{4}u^{5}+c_{5}u^{6}+c_{6}u^{7}+c_{7}%
u^{8}+c_{8}u^{9}+c_{9}u^{10}+c_{10}u^{11}+c_{11}u^{12}+c_{12}u^{13}+\ldots\\
\end{array}
\right)  .
\end{equation}
}
\end{widetext} 
Hence it follows%
\begin{small}
\begin{align*}
&a_{1}   =-\frac{2\lambda}{27},a_{2}=0,a_{3}=\frac{\lambda^{2}}{81}%
,a_{4}=0,a_{5}=\frac{\lambda^{3}}{162}, \\
& a_{6}=0,a_{7}=\frac{5\lambda^{4}}%
{1296},a_{8}=0,a_{9}=\frac{7\lambda^{5}}{2592}, \\
& a_{10}=0,a_{11}%
=\frac{7\lambda^{6}}{3456},a_{12}=0,\\
& b_{1}   =0,b_{2}=0,b_{3}=0,b_{4}=0,b_{5}=0,b_{6}=0, \\
& b_{7}=0,b_{8}%
=0,b_{9}=0,b_{10}=0,b_{11}=0,b_{12}=0,\\
&c_{1}   =-\frac{5\lambda}{54},c_{2}=0,c_{3}=-\frac{35\lambda^{2}}{648}%
,c_{4}=0,\\
& c_{5}=-\frac{35\lambda^{3}}{1296},c_{6}=0, c_{7}   =-\frac{175\lambda^{4}}{10368},\\
& c_{8}=0,c_{9}=-\frac{245\lambda^{5}%
}{20736},  c_{10}=0, \\
& c_{11}=-\frac{245\lambda^{6}}{27648},c_{12}=0.
\end{align*}
\end{small}
Therefore,
\begin{align*}
\alpha &  \mapsto\frac{2}{9}-\frac{\lambda u^{2}}{6}-\frac{\lambda^{2}u^{4}%
}{72}-\frac{\lambda^{3}u^{6}}{144}-\frac{5\lambda^{4}u^{8}}{1152} \\
& -\frac{7\lambda^{5}u^{10}}{2304}-\frac{7\lambda^{6}u^{12}}{3072}+\ldots,\\
\beta &  \mapsto\frac{1}{3}-\frac{\lambda u^{2}}{6}-\frac{\lambda^{2}u^{4}%
}{24}-\frac{\lambda^{3}u^{6}}{48}-\frac{5\lambda^{4}u^{8}}{384} \\
& -\frac{7\lambda^{5}u^{10}}{768}-\frac{7\lambda^{6}u^{12}}{1024}+\ldots,\\
\Omega_{m} &  \mapsto0,\\
x &  \mapsto u,
\end{align*}
and we have the parametrization%
\begin{align*}
\dot{\phi} &  =\sqrt{\frac{2}{3}}\theta u+\mathcal{O}(u)^{14},\\
\rho_{m} &  =-2\sqrt{6}\theta^{2}\lambda v_{2}e^{\phi/2}\sim\mathcal{O}(u)^{14},\\
\sigma &  =\theta\Big(  -\frac{7\lambda^{6}u^{12}}{1024}-\frac{7\lambda
^{5}u^{10}}{768}-\frac{5\lambda^{4}u^{8}}{384} \\
& -\frac{\lambda^{3}u^{6}}{48}
-\frac{\lambda^{2}u^{4}}{24}-\frac{\lambda u^{2}}{6}+\frac{1}{3}\Big)
+\mathcal{O}(u)^{14},\\
\mathit{E} &  =\theta^{2}\Big(  -\frac{7\lambda^{6}u^{12}}{3072}
-\frac{7\lambda^{5}u^{10}}{2304}-\frac{5\lambda^{4}u^{8}}{1152} \\
& -\frac
{\lambda^{3}u^{6}}{144}-\frac{\lambda^{2}u^{4}}{72}-\frac{\lambda u^{2}}{6}
+\frac{2}{9}\Big)  +\mathcal{O}(u)^{14},
\end{align*}
where we choose $\lambda v_{2}\geq0$.

The dynamics on the center manifold of the origin are dictated by a gradient-
like equation $u^{\prime}=-\nabla U(u)$. For $\lambda>0$, $\omega
=u\sqrt{\lambda}$, the equation transforms to
\begin{equation}
\scriptscriptstyle
\omega^{\prime}=-\frac{\omega^{15}\left(  \left(  7\left(  63\omega
^{6}+168\omega^{4}+352\omega^{2}+704\right)  \omega^{2}+10560\right)
\omega^{2}+33792\right)  }{1572864}%
\end{equation}
for which the origin is a degenerated minimum. For $\lambda<0$, $\omega
=u\sqrt{-\lambda}$, the equation transforms to
\begin{equation}
\scriptscriptstyle
\omega^{\prime}=\frac{\omega^{15}\left(  \left(  -7\left(  63\omega
^{6}-168\omega^{4}+352\omega^{2}-704\right)  \omega^{2}-10560\right)
\omega^{2}+33792\right)  }{1572864},
\end{equation}
for which the origin is a degenerated maximum. Therefore, for $\lambda>0$
(respectively, $\lambda<0$) the center manifold, and hence, the origin of the
system  is a local attractor (respectively, a saddle). In the original
variables mean that for $\lambda>0$ the past attractor of the dynamical system
is $P_{4}$, and for $\lambda<0$ is a saddle point. That is exactly the same
dynamics as for $P_{3}$. However, as we will see shortly, the physical
solution, although it is Bianchi I, has a different asymptotic expansion.

Now, we take the time reversal back and work in terms of $t$. Hence%
\begin{align*}
\dot{\theta}  &  =\theta^{2}\Big(  -\frac{147\lambda^{12}u^{24}}{524288}
-\frac{49\lambda^{11}u^{22}}{65536}-\frac{77\lambda^{10}u^{20}}{49152} \nonumber\\
& -\frac{77\lambda^{9}u^{18}}{24576}-\frac{55\lambda^{8}u^{16}}{8192}
-\frac{11\lambda^{7}u^{14}}{512}-1\Big) \nonumber\\
&  \sim-\theta^{2}-\frac{11}{512}u^{14}\left(  \theta^{2}\lambda^{7}\right)
+\mathcal{O}\left(  u^{16}\right)  ,
\end{align*}%
\begin{align*}
\dot{u}  &  =\frac{\theta}{3}\Big(  \frac{147\lambda^{12}u^{25}}{524288}
+\frac{49\lambda^{11}u^{23}}{65536}+\frac{77\lambda^{10}u^{21}}{49152} \\
& 
+\frac{77\lambda^{9}u^{19}}{24576}+\frac{55\lambda^{8}u^{17}}{8192}
+\frac{11\lambda^{7}u^{15}}{512}\Big) \nonumber\\
&  \sim\frac{11\left(  \theta\lambda^{7}\right)  u^{15}}{1536}+\mathcal{O}%
\left(  u^{16}\right)  .
\end{align*}
As before,
\begin{footnotesize}
\begin{subequations}
\begin{align}
&\theta(t)=-\frac{c_{3}}{c_{3}(t_{0}-t)+\varepsilon} \nonumber\\
&\sim\frac{1}{t-t_{0}}%
+\frac{\varepsilon}{c_{3}(t-t_{0})^{2}}+\frac{\varepsilon^{2}}{(t-t_{0}%
)^{3}c_{3}^{2}}+\mathcal{O}\left(  \varepsilon^{3}\right), \\
& u=c_{4}\varepsilon,
\end{align}
however for this point%
\end{subequations}
\begin{align}
\sigma &  =\frac{1}{3(t-t_{0})}+\frac{\varepsilon}{3(t-t_{0})^{2}c_{3}} \nonumber \\
& +\frac{\left(  \frac{2}{c_{3}^{2}}-(t-t_{0})^{2}\lambda c_{4}^{2}\right)
\varepsilon^{2}}{6(t-t_{0})^{3}}+\mathcal{O}\left(  \varepsilon^{3}\right)
,\\
\mathit{E}  &  =\frac{2}{9(t-t_{0})^{2}}+\frac{4\varepsilon}{9(t-t_{0}%
)^{3}c_{3}} \nonumber\\
& +\frac{\left(  \frac{4}{c_{3}^{2}}-(t-t_{0})^{2}\lambda c_{4}%
^{2}\right)  \varepsilon^{2}}{6(t-t_{0})^{4}}+\mathcal{O}\left(
\varepsilon^{3}\right)  ,\\
\dot{\phi}  &  =\frac{\sqrt{\frac{2}{3}}c_{4}\varepsilon}{t-t_{0}}+\frac
{\sqrt{\frac{2}{3}}c_{4}\varepsilon^{2}}{(t-t_{0})^{2}c_{3}}+\mathcal{O}%
\left(  \varepsilon^{3}\right)  ,\\
\phi &  =\sqrt{\frac{2}{3}}c_{4}\varepsilon\ln(t-t_{0})-\frac{\sqrt{\frac
{2}{3}}c_{4}\varepsilon^{2}}{c_{3}(t-t_{0})}+\mathcal{O}\left(  \varepsilon
^{3}\right)  ,\\
\rho_{m}  &  =\frac{2\sqrt{6}c_{4}^{14}K_{0}\varepsilon^{14}}{(t-t_{0})^{2}%
}+\mathcal{O}\left(  \varepsilon^{15}\right)  .
\end{align}
\end{footnotesize}
where $c_{3}$ and $c_{4}$ are integration constants, and we set $\lambda
v_{2}=-K_{0}\varepsilon^{14}$, for a positive constant $K_{0}$. As for $P_{3}%
$, for $\lambda>0$, $\theta(t)\rightarrow\frac{1}{t-t_{0}}$ as $t\rightarrow0$
($\tau\rightarrow-\infty$). Hence, $P_{4}$ it is associated with (an
anisotropic) initial singularity. However, as per the physical solution
referred to, this is a different solution with different asymptotic expansions
for $\sigma,\mathit{E}$.

\end{document}